\RequirePackage{fix-cm}
\documentclass[twocolumn]{svjour3}          % twocolumn
\smartqed  % flush right qed marks, e.g. at end of proof
\usepackage{graphicx}
\usepackage[usenames]{color}

\usepackage{nicefrac}

\usepackage{amsmath,amsfonts,amstext,bm}

\usepackage{subfigure,bm,amsmath,amstext,amssymb} 
\usepackage{multirow}
\usepackage{booktabs}

\usepackage[none]{hyphenat}

\journalname{CGI2019 - TVCJ} % The correct name will be entered by the editor
\begin{document}

\title{CrowdEst: A Method for Estimating (and not Simulating) Crowd Evacuation Parameters in Generic Environments}

\subtitle{}
\author{Est\^ev\~{a}o Testa, Rodrigo C. Barros, Soraia Raupp Musse}
\institute{Pontifical Catholic University of Rio Grande do Sul, School of Technology, Graduate Program in Computer Science, Porto Alegre, Brazil}
\date{ }% The correct dates will be entered by the editor

% \author{ Est\^ev\~{a}o Testa, Rodrigo C. Barros, Soraia Raupp Musse}
        
        %\thanks{Soraia Musse, Rodrigo C. Barros, and Est\^ev\~{a}o Testa are at Pontifical Catholic University of Rio Grande do Sul
%Corresponding author: soraia.musse@pucrs.br}}

\maketitle

\begin{abstract} 
\sloppy
Evacuation plans have been historically used as a safety measure for the construction of buildings. The existing crowd simulators require fully-modeled 3D environments and enough time to prepare and simulate scenarios, where the distribution and behavior of the crowd needs to be controlled. In addition, its population, routes or even doors and passages may change, so the 3D model and configurations have to be updated accordingly. %many simulations may be necessary in order to find out a good evacuation plan, changing routes and decisions. 
This is a time-consuming task that commonly has to be addressed within the crowd simulators. %,Since the amount of people in a given simulated scenario can change over time, several simulations are often required in order to generate an optimal evacuation plan. 
With that in mind, we present a novel approach to estimate the resulting data of a given evacuation scenario without actually simulating it. For such, we divide the environment into smaller modular rooms with different configurations, in a divide-and-conquer fashion. Next, we train an artificial neural network to estimate all required data regarding the evacuation of a single room. After collecting the estimated data from each room, we develop a heuristic capable of aggregating per-room information so the full environment can be properly evaluated. Our method presents an average error of $5\%$ when compared to evacuation time in a real-life environment. 
%\blue{, which is a fine result for a first attempt with the kind of abstraction used}. 
%Although one can think that is may be a higher error value, we argue that it is also the error margin when we compare some crowd simulations and real life experiments, which is acceptable in the area. 
Our crowd estimator approach has several advantages, such as not requiring to model the 3D environment, nor learning how to use and configure a crowd simulator, which means any user can easily use it. Furthermore, the computational time to estimate evacuation data (inference time) is virtually zero, which is much better even when compared to the best-case scenario in a real-time crowd simulator.

%Evacuation plans have been historically used as a safety measure for the construction of buildings. The existing simulators require fully-modeled 3D environments and enough time to prepare and simulate scenarios. In addition, since the amount of people in a given simulated scenario can change over time, several simulations are often required in order to generate an optimal evacuation plan. With that in mind, we present in this paper a novel approach to estimate the resulting data of a given evacuation scenario without actually simulating it. For such, we divide the environment into smaller modular rooms with different configurations, in a divide-and-conquer fashion. Next, we train an artificial neural network to estimate all required data regarding the evacuation of a single room. After collecting the estimated data from each room, we developed a  heuristic capable of aggregating per-room information so the full environment can be properly evaluated. Our method presents errors within the 30\% margin when compared to evacuation time in a real and complex environment (a nightclub). In addition, it is not necessary to model the 3D environment, learn how to use and configure a crowd simulator, and the computational time to estimate is instantaneous when compared to a best case real-time simulator. 
\end{abstract}




%\maketitle

\sloppy

\section{Introduction}
\label{sec:intro}
As new buildings are designed and constructed by engineers and architects, evacuation procedures to assure the needed safety standards are a major concern. Evacuation drills are usually used to analyze and evaluate predefined evacuation plans, but despite presenting strong similarities with real-world emergency scenarios \cite{pauls1980building}, they still pose significant ethical, practical, and financial challenges to be addressed \cite{gwynne1999review}. % TO DO: pegar mais descrição dessa referencia pra ficar bonitão

Crowd simulation is an interesting tool for evaluating real-world behavior of crowds in controlled scenarios. It can be defined as the process of simulating the movement of large amounts of agents, or crowds, in a previously-defined environment. The different ways in which crowds can behave has been object of research for almost thirty years \cite{survey2015}, including a variety of fields such as architecture, computer graphics, physics, robotics, safety engineering, training systems, psychology, and sociology \cite{survey2015,soraia2013crowd}.

% When simulating crowds, one should consider a set of parameters that coherently reproduce the desired scenario. Cassol et al.

% \cite{Cassol2017} categorized those parameters in the following taxonomy: 

% \begin{itemize}
% \item \textit{Physical Structure}: dimensions, number of floors, number of rooms, location of exits, and stairs; 
% \item \textit{Functionality}: whether people act as though they were in an office, hospital, school, airport, stadium, or arena; 
% \item \textit{Population}: number and spatial distribution of people in the environment, age, gender, relationships among them, knowledge about the environment; and 
% \item \textit{Event Conditions}: factors that may affect the navigability on the scenario, such as time (day or night), smoke, fire, or heat. 
% \end{itemize}

Regarding evacuation planning, crowd simulation can be used to evaluate evacuation plans, given a parameterized environment. For instance, Cassol et al.~\cite{Cassol2017} employed \textit{CrowdSim}~\cite{Galea:1998,cassol:2016}, which is a crowd simulator tested and validated in real-world scenarios, to create and gather data from simulations. They also employed CMA-ES~\cite{hansen1996cma-es,hansen2011cma-es}, an evolutionary algorithm that varies the population data so that different portions of the crowd follow different routes. %The simulation data was then evaluated using the $ep$ metric~\cite{cassol:2016} in order to find which route configuration was best-suited for evacuation (more details on the metric in~\cite{Cassol2017}). 

Albeit simple, when considering all  possible evacuation routes an environment can have, a very large amount of simulations have to be executed, growing exponentially as more details are added to that environment. Even though simulations can run in a similar time needed for real-world evacuation drills, it is still the most computationally-demanding step in such an evacuation planning approach. 

Hence, this paper presents a method whose goal is to estimate results of evacuation scenarios without the need of actually doing simulations in a complex environment within its run-time. For achieving that goal, we divide the environment to be analyzed into a set of smaller connected rooms. We hypothesize that by estimating the evacuation data on each room, we are capable of estimating the required information for the environment as a whole. For estimating per-room data, we make use of a machine learning approach, namely Artificial Neural Networks (ANN), which are previously trained on data collected from a number of simulation scenarios simulated beforehand. Furthermore, we define a proper heuristic to unify per-room data in complex environments that comprise a set of connected rooms. To the best of our knowledge, this is the first work to makes use of a hybrid machine learning and crowd simulation approach for the task of evacuation planning in generic environments and to compare results with real-life scenarios.
%To the best of our knowledge, the objective of most works that uses machine learning in the crowds area is to count and estimate the quantity or density of peoples over an image of a crowd, only a few attempts were made to use machine learning together with crowd simulation for the task of evacuation planning.

%during environment parameters and the evacuation data from the rooms it connects. In order to coherently estimate evacuation data we investigated the usage of machine learning and neural networks algorithms using data generated in crowd simulator as the input. %validated with \textit{CrowdSim}. 
% This estimation of crowd evacuation scenarios using ANN is the main contribution of this work, since it is a new approach for Crowd Behavior as far as we know. We did not find any other work that used machine learning in crowd simulation area with similar purposes.

%------------------------------

This paper is organized as follows. Section~\ref{sec:related} presents related work on crowd simulation and evacuation planning. Section~\ref{sec:dev} describes our new approach to estimating parameters for evacuation plans based on machine learning. Section~\ref{sec:results} reports the results of a detailed experimental analysis for validating our new approach, whereas in Section~\ref{sec:remarks} we discuss our findings and point to future work directions.
%Finally, in Section~\ref{sec:remarks} we will provide insight about what still needs to be developed and how we plan to do it.

\section{Related Work}
\label{sec:related}
The study and modeling of crowd traffic and their behavior, as well as the environment and evacuation characteristics is vital to the use of critical spaces \cite{Fruin:1971,ma2014pedestrian,polus1983pedestrian,schadschneider2011empirical}. %Despite the fact that we have not found any approach that addresses  in the literature, 
We briefly review some existing papers in areas related to the three mains aspects of this paper: crowd simulation, evacuation planning, and crowd learning. We are not exhaustive in our analysis since the majority of them are not focused in the exact same problem we address on this work, i.e., estimating crowd parameters instead of simulating it.

\subsection{Crowd simulation}
Several studies were proposed to elaborate ways of simulating crowds in egress scenarios.%Examples of crowd-simulation tools include the already-mentioned \textit{CrowdSim} \cite{Galea:1998,Cassol:2012,cassol:2016}, as well as SAFEgress~\cite{Chu:2014}, MIMOSA~\cite{Huangreal:2010}.%, and BioCrowds~\cite{Bicho:2009}.

\textit{SAFEgress} (Social Agent For Egress)~\cite{Chu:2014} is a force-based approach that models evacuating pedestrians which are able to make their actions according to their knowledge of the environment and their interactions with the social groups and neighboring crowds. MIMOSA: Mine Interior Model Of Smoke and Action \cite{Huangreal:2010} is a specific application of agent modeling within the context of a virtual underground coal mine, with a fire and smoke propagation model, and a human physiology and behavioral model.
%\textit{BioCrowds} \cite{Bicho:2009} is a biologically-motivated approach that simulates crowds using a space-colonization algorithm, recreating several real crowds behaviors such as collision avoidance, speed reduction effect, and lane formation.
The approaches described in~\cite{Ji:2013,Aik:2012,Fu:2014} make use of cellular automata to reproduce pedestrian behavior and exit selection using a least-effort cellular automaton algorithm, in which the motions and goals are probabilistic.

Van Den Berg et al. \cite{ORCA} propose the method \textit{Optimal Reciprocal Collision Avoidance} ($ORCA$), which is used for robots to avoid collisions with each other. The method searches for the optimal velocity for each agent to move so all agents move through the environment without colliding. For that, it predicts the future positions of other agents and prioritizes velocities that minimize the probability of them colliding, ensuring that the agents adopt the velocities that will result in the lowest number of collisions by the time it performed the predictions.

%\subsection{Evacuation planning}
Regarding evacuation planning, the work of Cassol et al. \cite{Cassol2017} makes use of crowd simulation (\textit{CrowdSim}) and evolutionary strategy (CMA-ES) to search for the best configuration of routes for evacuation within a given environment. 
Similarly, Garrett et al. \cite{garrett2006evacuation} employ Evolutionary Computation methods to evolve the placements of exits and other equipment in an effort to minimize the simulated evacuation time of the environment occupants. The simulation is made using an artificial potential-fields model in which exits attract agents and obstacles and other agents repel them. 

\subsection{Crowd learning}
%In the crowd learning area we did not find works directly related to our work. Currently, a great focus of the area is on the crowd counting problem, also called density estimation. Chan et al. \cite{learning:chan2008privacy} created a privacy-preserving system for estimating the size of inhomogeneous crowds in a video, by segmenting it using a dynamic texture motion model, the correspondence of each segmented region's features and the number of people in it is estimated with a Gaussian Process regression. 
% by segmenting it using a dynamic texture motion model and extracting the features of each segmented region, the correspondence of the features and the number of people is estimated with a Gausean Process regression.

Tripathi et al.~\cite{Tripathi2018} recently surveyed techniques that have been used in the context of crowd analysis and convolutional neural networks.
In particular, the crowd learning area is heavily focused on the crowd counting problem, also called density estimation.  %Chan et al.~\cite{learning:chan2008privacy} create a privacy-preserving system for estimating the size of inhomogeneous crowds in a video, by segmenting it using a dynamic texture motion model and by analyzing the correspondence of features between each segmented region. The number of people in a video is finally estimated with Gaussian Process regression. %RCB: está bem estranha esta explicação do método
%EST: reescrevi abaixo mas acho que vou ler o artigo de novo pra redescrever melhor
Chan et al.~\cite{learning:chan2008privacy} create a privacy-preserving system for estimating the size of inhomogeneous crowds in a video. By segmenting the video and analyzing features detected between frames and regions, the number of people is estimated using Gaussian Process regression.
Similarly, Fradi et al.~\cite{learning:fradi2013crowd} extract features from videos and use a Gaussian symmetric kernel function to generate a crowd density map, allowing more specific locations of potentially-crowded areas. 

Liu et al.~\cite{mubassir2017} is the closest approach to our work. They make use of artificial neural networks to learn the behavior of simulated crowds with the objective of replacing simulation with estimations, which is basically our same goal. However, the authors simulate the crowd on a fixed environment (walls and routes) containing mobile parts whose disposition varies on each simulation. The position and rotation of these mobiles parts are used as input to the neural network to obtain the statistics of speed, number of collisions, and traveling time of each agent for this environment and crowd configuration. 

In this work we propose a methodology for estimating the evacuation of crowds in generic modular rooms using neural networks. These rooms can be connected to form larger and more complex environments, and we also propose heuristics to estimate this environment using the data obtained from estimating each particular room. %We make tests using different neural networks models and calculated the accuracy obtained for four estimated parameters.

\section{Proposed Approach}
\label{sec:dev}
In this section, we describe in details our proposed approach. First, we present the development of the crowd simulator that is employed to generate data for the learning process. Then, we present the process that was used to validate that simulator. Finally, we describe our per-room estimation strategy as well as the heuristics to unify the estimated results.

\subsection{Crowd Simulator}

To develop the crowd simulator, we make use of the \textit{Unity3D} game engine as a platform and implemented agents endowed with the main inherent collective behaviors as referred to in the literature: \textit{i)} \textit{goal seeking}, \textit{ii)} \textit{collision avoidance}, and \textit{iii)} \textit{least effort strategy} behaviors \cite{Fruin:1971b,henderson1971statistics,henderson1972sexual,henderson1974fluid,helbing:1997,Still:2000}.

A \textit{goal seeking} agent means that the agent will move inside the environment in order to reach a goal it is seeking. This behavior is simple to implement, just by keeping a goal position for each agent and moving the agent towards it.

When applying a \textit{least effort strategy}, the agent will move in trajectories that require less effort, avoiding turns while selecting the shortest paths. 
%This behavior may conflict with the \textit{collision avoidance}, since in most cases it is needed to change its orientation to avoid to collide with others. 
This behavior was implemented using path finding over a \textit{NavMesh}, a network of connected 3D planes which represents the navigable area. The \textit{NavMesh} used was the native's from \textit{Unity3D}.

When applying a \textit{collision avoidance} behavior, the agent will move avoiding physical contact with obstacles and with other agents. %In part, this behavior is implemented using the \textit{NavMesh} to avoid collision with obstacles. 
In order to do that, we implemented the \textit{ORCA} model \cite{ORCA}. The main goal of this method is to find the optimal velocity of each agent so that it manages to move through the environment without colliding with obstacles or agents. It corrects the course of their movements considering every other agent's velocity and position in a future time, reacting preemptively and avoiding possible collisions between them. We used as the optimization velocity for~\textit{ORCA} the direction each agent is following toward its goal on the \textit{NavMesh}. %An example of this collision avoidance can be seen in Figure~\ref{fig:colavoidance}.
% \begin{figure}[htb]
% \centering
%   \subfigure[Agents move in straight line toward their goals. Paths to goals in red.]{\label{fig:path1}\includegraphics[width=0.46\linewidth]{Figs2/agentePathfinding1}}
%  \quad
%  \subfigure[Agents change their paths to avoid collision between them.]{\label{fig:path2}\includegraphics[width=0.46\linewidth]{Figs2/agentePathfinding2}}
%  \subfigure[Agents resume their path toward their goals.]{\label{fig:path3}\includegraphics[width=0.46\linewidth]{Figs2/agentePathfinding3}}
% \caption{Example of \textit{collision avoidance} behavior in a simple environment. (a) Agents move towards their goals in a simple environment. (b) Agents avoiding collision. (c) Agents resume their paths.}     
% \label{fig:colavoidance}
% \end{figure}
%Lastly, we designed simulations to be used to train ANNs. We simulated rooms with 6 following parameters: \textit{Room width}, \textit{Room length}, \textit{Exit size}, \textit{Input flow}, \textit{Flow Duration} and \textit{Initial population}. Also, for each executed simulation we collected the following 
%data needed for calculating the \textit{EP} value parameters: (\textit{Total evacuation time},\textit{Average exit time},\textit{Average speed} and \textit{Average density}). This is going to be detailed in Section \ref{sec:database}. 
%As our simulator is going to be used to train and test the ANNs, we considered very important to evaluate its correctness. So, in next sections we describe the validations performed with our simulator.

%SO: tenho dúvidas sobre a próxima section no que tange o paper a ser submetido...
\subsubsection{Model validation}

% The validation of our crowd simulation was separated in two steps: evaluate our $ORCA$ implementation when compared to the original one, and validate its similarity with real crowds. 

\begin{figure}[htb]
\centering
 \subfigure{\label{fig:orcacase12}\includegraphics[width=0.47\linewidth]{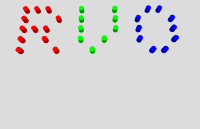}}
 \quad
\subfigure{\label{fig:orcaval12}\includegraphics[width=0.47\linewidth]{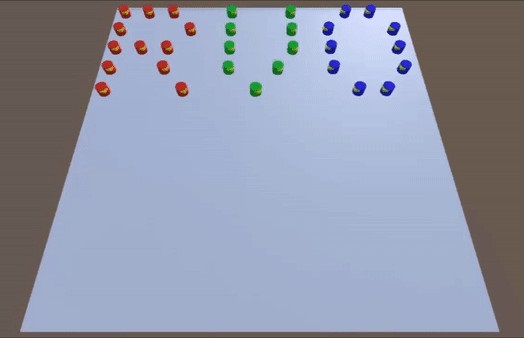}}

 \subfigure{\label{fig:orcacase32}\includegraphics[width=0.47\linewidth]{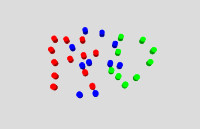}}
 \quad
\subfigure{\label{fig:orcaval32}\includegraphics[width=0.47\linewidth]{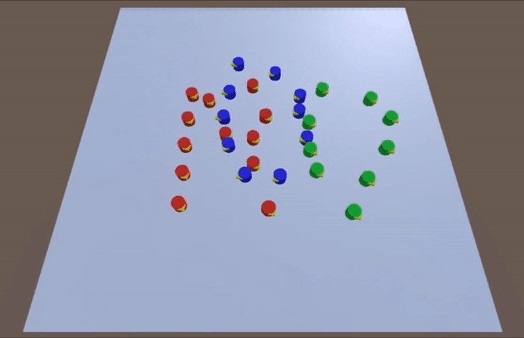}}

 \subfigure{\label{fig:orcacase52}\includegraphics[width=0.47\linewidth]{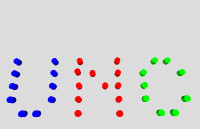}}
 \quad
\subfigure{\label{fig:orcaval52}\includegraphics[width=0.47\linewidth]{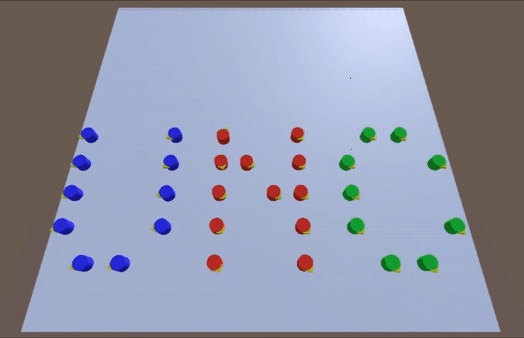}}

\caption{Comparison between the \textit{ORCA}'s showcase (left) and our simulation (right). A total of 41 agents are positioned to form the word ``RVO", and then move down the screen to form the word ``UNC" while avoiding collisions with each other.}     
\label{fig:showcase2}

\end{figure}

% For the first step the main behavior is the collision avoidance.% and it is illustrated in Figure~\ref{fig:colavoidance}. 
For validating our crowd simulation, we compare our approach with the original ORCA method. For doing so, we recreated one of the~\textit{ORCA}'s showcases available at the developer's website~\cite{ORCAsite}. Some images of this showcase can be seen in Figure~\ref{fig:showcase2}. 
% For the real crowd validation we used the same cases used to validate \textit{CrowdSim}~\cite{Cassol:2015} which was based on the \textit{guidelines for evacuation analysis for new and existing passenger ships} proposed by the International Maritime Organization (IMO)\cite{IMO:2007}. This validation is composed of component and qualitative tests. Further details are presented in Section~\ref{sec:IMO}.
%SO: section abaixo diminuir bastante para o paper a ser submetido
%\subsubsection{Comparing with \textit{ORCA} showcase}

One of the showcases was created to depict the agents' behaviors using \textit{ORCA} to avoid collision and its capacity to move efficiently while looking natural to the human eye. We recreated this showcase using solely visual information from real ORCA cases, and without knowing the agents parameters (e.g., agents positions or speeds), so we expect similar results from our model, though maybe not the same. % but maybe not the same. Nevertheless, a good visual comparison of the behaviors could still be done.

%The first showcase consists of an environment with no obstacles and ten agents positioned in a circular pattern, in which each agent is supposed to move towards the opposing end of the circle, while avoiding collision with the others. As the agents move towards the center, the space between them becomes smaller and some of them will need to assume non-optimal routes and reduce velocities to avoid the collisions. In both implementations (ORCA showcase and ours), agents move successfully towards their goals, reduce their speed and change their courses to avoid collision with others. Both the first showcase and our simulation can be visualized in Figure~\ref{fig:showcase}.

% \begin{figure}[htb]
% \centering
%  \subfigure{\label{fig:orcacase1}\includegraphics[width=0.45\linewidth]{Figs2/validateCase1}}
%  \quad
% \subfigure{\label{fig:orcaval1}\includegraphics[width=0.45\linewidth]{Figs2/validatedCase1}}

%  \subfigure{\label{fig:orcacase3}\includegraphics[width=0.45\linewidth]{Figs2/validateCase3}}
%  \quad
% \subfigure{\label{fig:orcaval3}\includegraphics[width=0.45\linewidth]{Figs2/validatedCase4}}

%  \subfigure{\label{fig:orcacase5}\includegraphics[width=0.45\linewidth]{Figs2/validateCase5}}
%  \quad
% \subfigure{\label{fig:orcaval5}\includegraphics[width=0.45\linewidth]{Figs2/validatedCase5}}

% \caption{Comparison between the \textit{ORCA}'s showcase (left) and our simulation (right). Agents are positioned in a circular pattern and moves to the opposite end of the circle while avoiding collisions with others.}     
% \label{fig:showcase}

% \end{figure}

The showcase consists of 41 agents organized into three groups, where the agents from each group are positioned to form the letters of the word ``RVO". Then, they proceed to move down the screen to form the word ``UNC". The group forming the letter ``R" moves to form the letter ``N" at the end of the showcase, whereas the one forming the letter ``V" goes on to form the letter ``C". Similarly, the one forming the letter ``O" follows to forms the letter ``U". Easier movements could be made to complete this transformation, however this specific movement was chosen in order to promote more potential collisions, thus requiring the use of collision avoidance mechanisms, which is ideal for testing ORCA. A comparison showing the showcase and our simulation as well as the movement pattern generated by both simulators can be visualized in Figure~\ref{fig:showcase2}. In both cases, the agents succeeded in moving and forming the other word, in a quite similar way. %Visually comparing them we say the agents in the showcase had a little more scattered movement pattern, this may be due to the difference in the size of the environment or the difference in the prediction time ORCA uses to make the agents react to the possible collisions. Please, notice, that as we said before we did not know the used parameters, but just imitating the showcases based on visual results.

\subsection{Estimation of Rooms Parameters based on ANNs}
\label{sec:room_estimation}

%Once last sections validate our simulator, t
This section presents our model to estimate crowd data resulting from the evacuation process instead of having to simulate it. We are interested on testing Artificial Neural Networks as a tool to estimate crowd parameters instead of generating them through simulations. The overview of our method comprises three phases: i)~performing several simulations in order to generate the training dataset for the ANN (see Section~\ref{sec:database}); ii)~ANN training and validation, as discussed in Section~\ref{sec:trainingmodel}; iii)~testing the learning methodology for rooms estimation in complex environments (Section~\ref{sec:env_estimator}).
%we The ANN is trained using simulations performed with our simulator w.r.t various configurations of rooms and population, as following described: 
%SuAs a summary of the method we designed for the estimations, it basically sees the environment as a set of modular connected rooms, each room containing its own values of:

The room $i$ parameters are described as follows:

\begin{itemize}
\item $Width_i$ is the room width (meters);
\item $Length_i$ is the room length (meters); 
\item $es_i$ is the exit size (meters); 
\item $f_i$ is the input flow: agents per second that enter within the room;
\item $F_i$ is the flow duration: duration in seconds defining the period where agents enter the room;
\item $ip_i$ is the initial population: number of agents which are inside the room at the beginning of the simulation.
\end{itemize}

Next, the estimation proceeds by estimating the \textit{evacuation total time} ($tt_i$), in seconds, for the $i^{th}$ room. Details on the estimation procedure is given in the following sections.

% \begin{itemize}
% \item \textit{Evacuation total time} $tt_i$ (seconds), 
% \item \textit{Average exit time} $\bar{t_i}$ (seconds), 
% \item \textit{Average speed} $\bar{s_i}$ (meters per second) and 
% \item\textit{Average density} $\bar{d_i}$ (agents per meter).
% \end{itemize}

% Then the estimation proceeds by estimating the following data for each room $i$:

% \begin{itemize}
% \item \textit{Evacuation total time} $tt_i$ (seconds), 
% \item \textit{Average exit time} $\bar{t_i}$ (seconds), 
% \item \textit{Average speed} $\bar{s_i}$ (meters per second) and 
% \item\textit{Average density} $\bar{d_i}$ (agents per meter).
% \end{itemize}
 
%After the data of all rooms were estimated it proceeds by estimating this same data but for the entire environment. This modularization is done to allow us to estimate different environments with different layouts while maintaining the same estimation procedure. 

\subsubsection{Dataset Creation\label{sec:database}}

To create a dataset of simulations that could be used to train and test the learning algorithms, we automatize the process of creating rooms. This process generates simulations based on the input parameters, which are randomized for each simulation, and it collects data of the output parameters. Each generated simulation comprises a rectangular room with a single exit. In this paper, we use specifically the scope of rectangular rooms to represent the walkable space in a certain room and single exits in the training dataset. We decided to do so because if we were to include all possible geometries for rooms and all possible exit numbers, this would result in a combinatorial explosion of  environment possibilities to be simulated.

Note that during inference time we can test non-rectangular rooms with more than one exit. For such, we have to approximate the room by a rectangle, and if we want to consider rooms with more than one exit, the user has to inform a single exit whose size is the sum of the size of all available doors.  %represent the the free space to exit the room. } %representing the size we simply divide the room uniformly, as the population, in as many rooms as are the number of exits.} %Anyway, other rooms shapes should be addressed in a future work
The room's dimensions vary according to the \textit{room width} and \textit{room length} parameters, whereas the exit size varies according to the \textit{exit size} parameters. 

\begin{figure}[htb]
\centering
\includegraphics[width=0.75\linewidth]{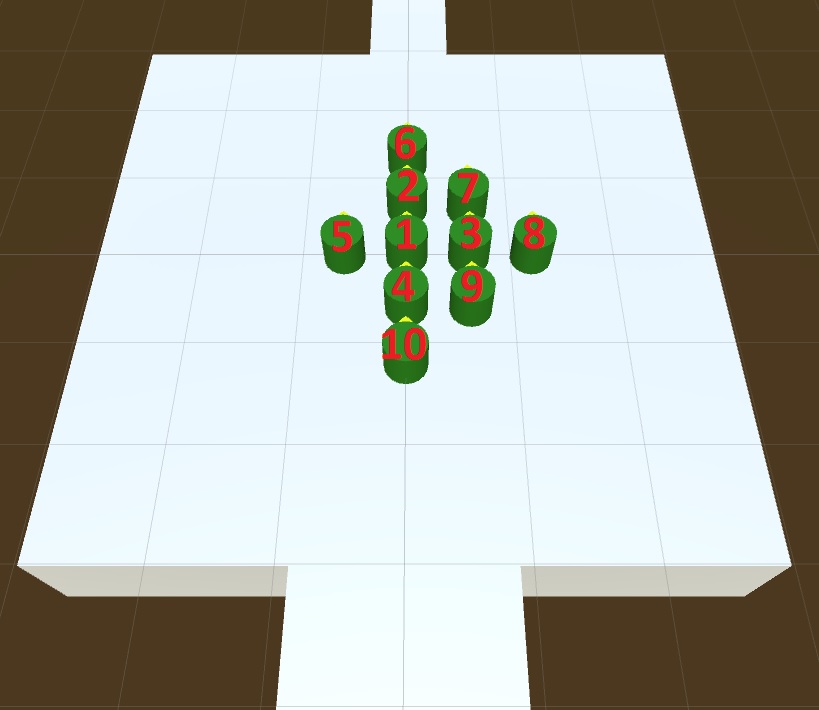}
\caption{Agents placed in a squared spiral pattern in a $6\times 6$ room in the crowd simulator, numbered according to the order they are placed.}     
\label{fig:diamond}
\end{figure}

Agents are placed in the room using the \textit{initial population}, \textit{input flow}  and \textit{flow duration}  parameters. The agents informed in the initial population are those already in the room at the beginning of the simulation.
They are created and placed in a squared spiral pattern, forming a diamond at the center of the room as more agents are placed, to avoid agents overlapping each other. Figure~\ref{fig:diamond} shows agents placed according to this pattern. As the number of agents rises, the area of the diamond increases. However, since their positions are restricted by the size of the room, if no space is available to accommodate all agents that are created it may be inevitable that some agents will start with overlapping positions. If this happens, agents will try to avoid collisions among them when the simulation starts, eventually finding enough space to move without colliding as other agents exit the room.

New $n$ agents are created w.r.t to the flow information. These agents are created at the entrance in the opposite direction of the exit in room $i$, considering a constant flow defined as: 

\begin{eqnarray}
n_i = f_i \times F_i,
\end{eqnarray}
where $f_i$ is the \textit{input flow} defined for room $i$, and
$F_i$ is the \textit{flow duration} for $r_i$. Indeed, we consider that the input flow in a certain room to be simulated is constant because we do not have more detailed information considering that we are only simulating a single room. If crowds are self-organized, which is what happens when people have space and time to adapt to the restricted physical space, their flow tend to be constant (e.g., already organized in a previous room in the environment hierarchy)~\cite{Helbing:2005} and~\cite{Musse:2012}.

%at a time $t$ that is included in the interval $[t_i;t_f]$ and represents respectively initial and final time for the people flow in the room.
%When $t$ is greater than $t_f$, i.e. the people flow in room $i$ is finished, $F_i^t=0$, so the entrance of agents in room $i$ stops. 

%Agents always enter into the room from an entrance in the opposite side of the exit.
%SO: frase estranha aqui abaixo | ajustei um pouco
%This entrance size is proportional to the \textit{Input flow}, so that it has necessary space to continuously place new agents without overlapping the ones that were just created moments ago.
%SO: Frase estranha acima... entrance size (meters) é proporcional ao flow? que proporção desejada é essa? Acho melhor comentar a frase acima... Já falamos mais acima que pode ter overlaps temporários... | ok, then ajustei tudo pra ficar direito

Regarding the remaining parameters present the simulation: the space each agent occupies is represented through a cylinder with a radius of 0.3 meters and it moves trying to maintain a max desired speed of 1.2 m/s. Both of those values are extracted from the literature %for the average value people size and the average speed people have while walking 
\cite{tilley2002measure}.

%\textit{ORCA} makes use of a prediction time to make agents react and avoid collisions with each other. We were not able to find any research or study about which value must be used to be more close to reality. In fact, the common assumption is that value can vary according to each individual (cultural or personality) and with the environment conditions such as time of day, smoke, fire and crowd density. 
%\blue{In our simulations we empirically adopted the value of 3 seconds because it was the value which we got better visually convincing collisions avoided during simple tests. When we used prediction time equal to 1 second, the method provided a strange behavior when one agent passes by another one in low densities. In this case, they got too close to each other, while having free space to navigate around. When we used 5 seconds the agents became too much generous, frequently stopping to allow others to pass even from far away distances.}

%SO: fiz uma pequena alteração, veja o que achas.

To generate the training and validation sets, the parameters of the room were randomly defined within the intervals specified in Table~\ref{tab:values}.
We generate 18000 rooms for training and 2000 for validation. The tests are presented later in Section~\ref{sec:results}.
\begin{table}[h]
\centering
\begin{tabular}{|c|c|c|c|}
\hline
Parameter & Min value & Max value & Value Type \\ \hline
 \textit{Room width}  & 2.0 & 20.0 & Float        \\ 
  ($width_i$) &  &  &         \\ \hline
\textit{Room length}  & 2.0 & 20.0 & Float        \\
 ($length_i$) &  &  &         \\\hline
\textit{Exit size}  & 0.9 & 5.0 & Float \\ 
 ($es_i$) &  &  &  \\
\hline
\textit{Input flow}  & 1.0 & 10.0 & Float \\
 ($f_i$) &  &  &  \\
\hline
\textit{Flow duration}  & 0.2 & 100.0 & Float
\\
 ($F_i$) &  &  & 
\\
\hline
\textit{Initial population} & 0 & 99 & Integer
\\
$ip_i$ &  &  & 
\\
\hline
\end{tabular}

\quad

\caption{Intervals for each parameter in the dataset generation process.}
\label{tab:values}
\end{table}

%SO: For 
%The same was made for the testing case, however the case created in the Environment Editor was modeled after a real world case with a real scenario.

\subsubsection{Training and validation\label{sec:trainingmodel}}

We train an ANN  to estimate the evacuation total time $tt_i$.  %, Average exit time, Average speed and Average density) we trained a different neural network, this was done to avoid connecting the learning process of one parameter to the others, as it is possible that a network model works best for one while other model is best for other and that one training could better identify Evacuation Time over Average density and another training the opposite, for example.}
The training process was performed using the Stochastic Gradient Descent (SGD) optimization approach for 100 epochs with a $10e-7$ learning rate, stopping only when no more reduction on the loss was observed in new epochs even when reducing the learning rate.
%\blue{For each attribute to be estimated (Evacuation time, Average exit time, Average speed and Average density) we trained a different neural network, this was done to avoid connecting the learning process of one parameter to the others, as it is possible that a network model works best for one while other model is best for other and that one training could better identify Evacuation Time over Average density and another training the opposite, for example.}
%We started with a simple model of neural network: one hidden layer with six neurons fully connected and no bias. The loss function used was the mean squared error function. 
%\blue{The training process was made using Stochastic Gradient Descent (SGD), stopping only when no more reduction on the loss was observed by new epochs even when reducing the learning rate.}
%\blue{ Vi os tipos de gradiente descendente que tem e na real a gente fez pelo stochastic. }
Once the $tt_i$ is estimated, the validation errors were measured using the absolute relative error comparing the predicted value and the simulated value, here considered as the ground truth. We measured the quality of our estimations according to how many cases had less than 10\% error.

\begin{figure}[htb]
\centering
\includegraphics[width=1\linewidth]{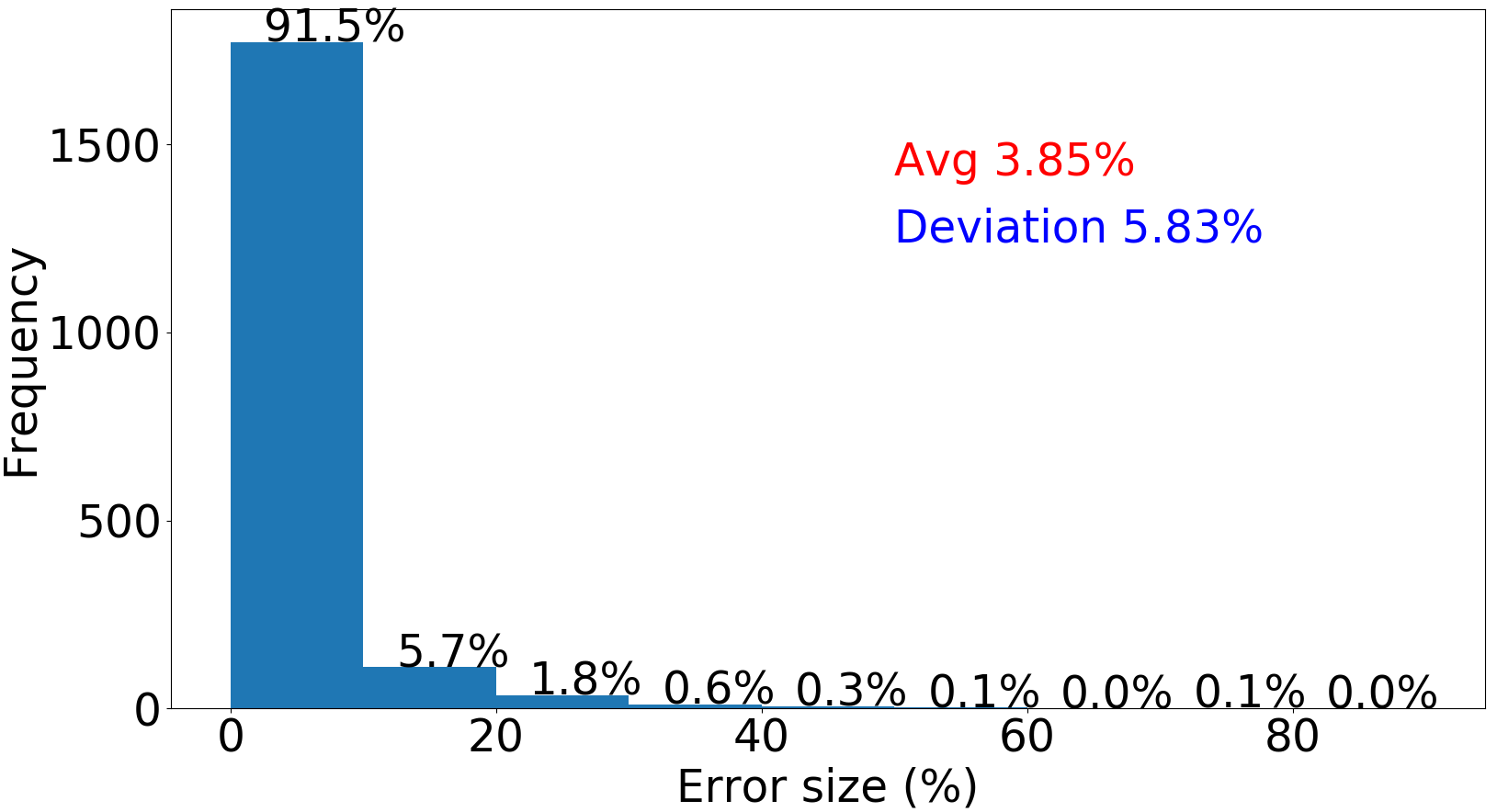}
\caption{Hidden layer with 400 neurons, total evacuation time error percentage.}     
\label{fig:ST400neurons}
\end{figure}

We experiment with 1, 2, and 3 hidden layers with the 6 values of input from the dataset, 6 neurons on each layer and a single output value: the evacuation total time estimated. From those models, the one which achieved the best accuracy was the one with the single hidden layer. Then we proceeded to test several number of neurons on this layer. We tested with 2, 3, 6, 50, 200, 400 and 500 neurons on the hidden layer and we got the best validation accuracy with 400 neurons. Then, we proceeded to train the defined ANN stopping only when no more reduction on the loss was observed in new epochs, even when reducing the learning rate. The evacuation time in the 2000 rooms estimated by a 400 neurons ANN resulted in 91.5\% of the validation cases below 10\% error, as illustrated in Figure~\ref{fig:ST400neurons}.

The next section describes our heuristic capable of aggregating per-room information in order to estimate complex environments.

% \begin{figure}[htb]
% \centering
% \includegraphics[width=1\linewidth]{Figs/scattertraining.png}
% \caption{Search for a better network model. Attempts with 6, 50, 200, 400 and 500. Plotted number of cases bellow 10\% error in the individual validation.}     
% \label{fig:treinamentos}
% \end{figure}

% \begin{figure}[htb]
% \centering
% \includegraphics[width=1\linewidth]{Figs/400_neurons_-Average_exit_time__seconds_.png}
% \caption{Hidden layer with 400 neurons, average exit time error percentage.}     
% \label{fig:AE400neurons}
% \end{figure}

% \begin{figure}[htb]
% \centering
% \includegraphics[width=1\linewidth]{Figs/200_neurons_-Average_speed__m_s_.png}
% \caption{Hidden layer with 200 neurons, average speed error percentage.}     
% \label{fig:AS200neurons}
% \end{figure}

% \begin{figure}[htb]
% \centering
% \includegraphics[width=1\linewidth]{Figs/200_neurons_-Average_density.png}
% \caption{Hidden layer with 200 neurons,average density error percentage.}     
% \label{fig:AD200neurons}
% \end{figure}

%SO: aqui abaixo acho que vai apra results

% Validating these new networks with the environments cases, figure \ref{fig:400-400-200-200}, improvements in the Simulation time, Average speed and Average density can be observed, from 22\% average error to 19\%, 61\% to 52\% and 53\% to 46\% respectively, while maintaining basically the same standard deviation. Average exit time suffered from 65\% to 71\%, keeping the same standard deviation.

% \begin{figure}[htb]
% \centering
% \includegraphics[width=1\linewidth]{Figs/400-400-200-200_average.png}
% \caption{Validation of estimated environments with 14 environment cases using networks with 400 hidden neurons for Simulation time and Average exit time and 200 neurons for Average speed and Average density. The average error and standard deviation of each attribute is shown.}     
% \label{fig:400-400-200-200}
% \end{figure}

%AQUI

\subsection{Environment Estimation based on ANNs and Heuristics}
\label{sec:env_estimator}

As described in last section, we are now capable of estimating the time of evacuation for each room in the environment. %, four of them focused on learning specific crowd parameters, i.e. Simulation time, Average exit time, Average speed and Average density. 
However, crowd evacuation methods are used in general for more complex environments than a single room. Therefore, 3D modeling is usually necessary in order to create the environment to be simulated. Since we are proposing crowd estimation instead of simulation, we propose to create a simple environment based on a graph of rooms (i.e., an environment graph where the connections among the rooms are defined as graph edges). We developed an environment editor (see Section~\ref{sec:EnvEditor}), where the user can easily create and edit a graph representing an environment to be estimated. 

Our proposal is to use ANNs to estimate the parameters from each room (as discussed in the last section) and then combine this data via empirically defined heuristics to compute the global data for the entire environment. %Heuristics are used in two phases. The first one is responsible for considering the hierarchy of rooms adjusting the population and flows of rooms that lead people to other rooms. Then, in a second step, the global data regarding the whole environment is computed.
%For each data to be estimated in the room a neural network is trained, it receives the values of the rooms as input to estimate the resulting data of the room. 
Environment $e$ comprises $N$ rooms $r$ to be estimated. We estimate the crowd parameters of each room $r_i$ using the ANN for obtaining $tt_i$. Therefore, we have two types of rooms within an environment $e$: i)~rooms of type $D$, whose population impacts another room, i.e., people from $r_{D_i}$ goes to another room; and ii)~rooms of type $E$ (exit rooms), which are rooms that lead the population to the output of the environment. Moreover, those two types of rooms are mutually-exclusive, i.e., rooms $D$ never lead to exits and rooms $E$ never impact any other room in $e$. %The details about the training of those neural networks are present on chapter \ref{sec:results}.

\begin{figure*}[htb]
\centering
\includegraphics[width=1\linewidth]{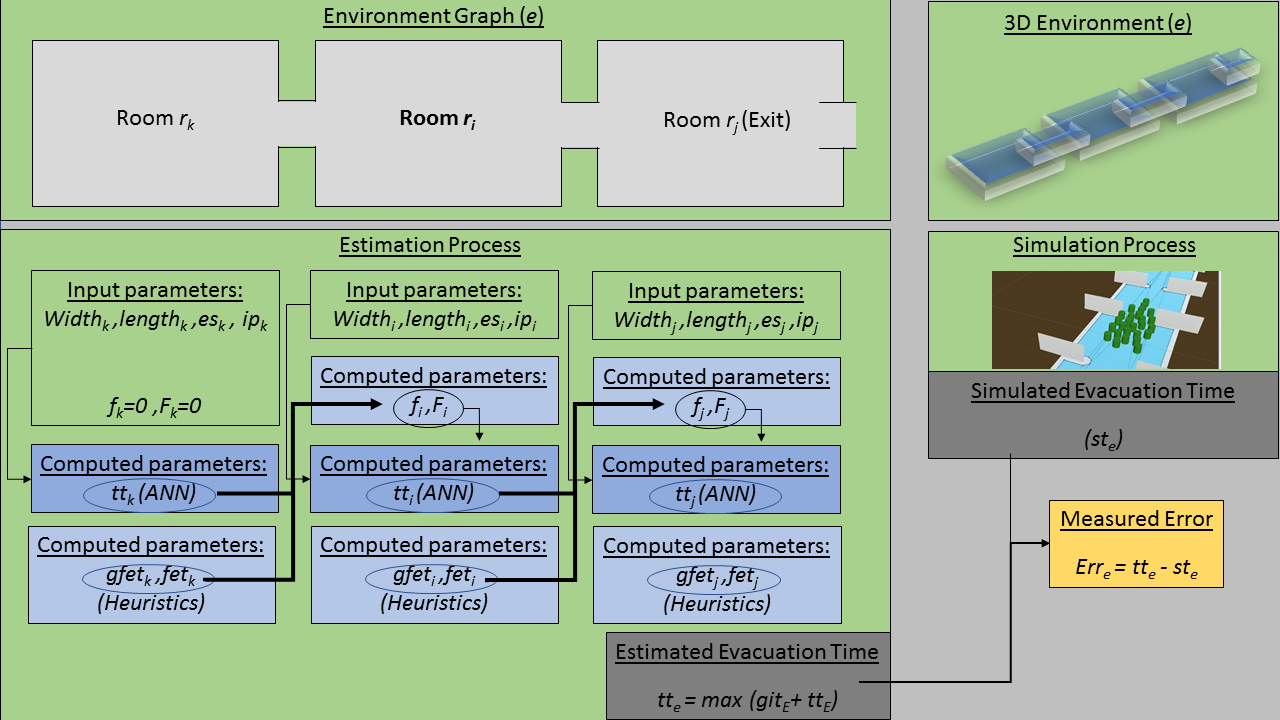}
\caption{Overview of the estimation method. The environment graph shows the hierarchy of rooms to be considered and how they affect other rooms. On the right we have the same environment but modelled in 3D and simulated using our crowd simulation. In addition, the measured error is also illustrated.}
\label{fig:Heuristics}
\end{figure*}

Figure~\ref{fig:Heuristics} presents the overview of our estimation method. Note that some parameters are input for the rooms structure. For instance, $r_k$ (room  without dependent rooms, on the left) has the following input data: $width_k$, $length_k$, $es_k$ (exit size) and $ip_k$ (initial population). In addition, for this specific room the input flow $f_k=0$ and flow duration $F_k=0$ because no agents arrive in such space coming from other rooms. These are the input parameters for running the ANN and it results in $tt_k$ (total time of simulation of $r_k$). In addition, variables $gfet_k$ (global first exit time) and $fet_k$ (local first exit time) are computed using our heuristics (described later) and impact the next room in the hierarchy. $r_i$ is a room with a dependent room ($k$) and it is also a dependent room regarding room $j$. Because $r_i$ has a dependent room, $f_i$ and $F_i$ are no longer $0$ and are impacted by $r_k$ data, as shown in Figure~\ref{fig:Heuristics}. It means that variables $gfet_k$ and $fet_k$ are used to compute $f_i$  and $F_i$, which are then used as input to the ANN together with the remaining input parameters: $Width_i$, $length_i$, $es_i$ and  $ip_i$ to estimate $tt_i$. The same process happens for room $j$. On the right side of the figure, we illustrate the simulation process that happens using the 3D environment, and finally the measured error can evaluate the correctness of our estimation process ($Err_e$), where $e$ is the simulated and estimated environment.

\vspace{0.5cm}
\noindent\textit{A~-~ Heuristics to Estimate per-Room Data}
\vspace{0.5cm}

The environment estimator loads the specified environment (using the editor presented  in Section~\ref{sec:EnvEditor}) and performs the data estimation using the ANN for each room separately. % (types $D$ and $E$). %using the two specific ANNs (as described in Section~\ref{sec:room_estimation}). 
%Obtained data for room $r_i$ are: \textit{Evacuation total time} ($tt_i$) and \textit{Average exit time} ($\bar{t_i}$).
%\textit{Average speed} ($\bar{s_i}$) and \textit{Average density} ($\bar{d_i}$). 
The ANN outputs the estimated total time ($tt$) for each room. However, some rooms are dependent on others, i.e., a given room $r_k$ (as illustrated in Figure~\ref{fig:Heuristics}) is impacted by other rooms (type $D$) whose population moves towards $r_k$. This is the definition of the set $r_{D_k}$ of dependence rooms that impact $r_k$. That is why some rooms have to be estimated before others. We deal with this problem by computing global and local-time parameters in order to synchronize the rooms. First we are going to present the execution flow for rooms that do not have dependence ($N_{D_k}=0$), i.e., the first rooms within the environment graph. 

%The execution flow for each room $r_i$ is following described in five steps:
\begin{enumerate}
%\item {To compute $D_i$, i.e. the dependence rooms of $r_i$};
\item Since nobody enters room $k$ coming from another room (as illustrated in Figure~\ref{fig:Heuristics}), then the \textit{input flow} $f_k=0$ and the \textit{flow Duration} $F_k=0$.  
%\item \textit{Initial population} $pop_i = ip_i$, % ip_i só é mencionado depois

\item Once all parameters for $r_k$ are set ($width_k$, $length_k$, $es_k$, $ip_k$ --- those four coming from the Environment Editor --- and $f_{k}=0$ and $F_{k}=0$), the estimation of $r_{k}$ can happen normally. The neural network is loaded and the total time is estimated ($tt_{k}$). % and  $\bar{t_{i}}$) %, $\bar{s_{i}}$ and $\bar{d_{i}}$) 
%are estimated using their respective networks. 

\item In order to provide data regarding the agents that leave $r_k$ and go to the next room, two other parameters are computed, $fet_k$ and $gfet_k$, as follows:

%\begin{eqnarray}
 %fet_{i} = max(0,\bar{t_{i}} - \frac{(tt_{i} - \bar{t_{i}})}{\delta}), 
 %fet_{i} = \frac{length_{i}}{maxSpd}
 %\label{eq:fet}
 %\end{eqnarray}
 
 %\begin{eqnarray}
 %fet_{i} = \frac{\frac{length_{i}}{2} - Max(0,(\frac{0.6 * \sqrt{ip_{i} * 2 - 1}}{2})}{maxSpd} ,
%\label{eq:fet2}
%\end{eqnarray}

%\begin{eqnarray}
% fet_{k} = 
% \begin{cases}
%     \frac{length_{k}}{maxSpd},& ip_{k} = 0 \\
%     \frac{\frac{length_{k}}{2} - Max(0,(\frac{0.6  \sqrt{ip_{i} . (2 - 1)}}{2})}{maxSpd},              & ip_{i} > 0
% \end{cases}
% \label{eq:fet}
% \end{eqnarray}

 \begin{eqnarray}
 fet_{k} = 
 \begin{cases}
     0, & ip_{k} = 0 \\
     %\frac{length_k}{maxSpd}, & ip_{k} > 0
     \frac{\frac{length_k}{2}}{MaxSpd}, & ip_k > 0,
 \end{cases}
 \label{eq:fet}
 \end{eqnarray}
where $fet_{i}$ is the first exit time from $r_k$, meaning the time that the first agent exits that room, and $ip_{k}$ is the number of agents initially created in room $k$. $fet_{i}=0$ if there are no agents inside $r_k$. Conversely, %it is simply calculated as the time it takes for an agent to pass through the room at maximum speed as defined in the simulator ($1.2m/s$). However, when 
if $ip_{k} > 0$, the first agent to exit will not need to walk throughout all the room length, so we consider half of the room length as an average for all agents to walk and exit the room. %. To deal with that we designed the second part of~Equation~\ref{eq:fet}, which represents the distance reduction  to be traveled by the first agent to exit the room. 
The next equation is computed afterwards the ANN execution:

\begin{eqnarray}
 gfet_i = fet_i,
\label{eq:it}
\end{eqnarray}
where $gfet_i$ is the time the first agent exited the room considering all rooms. For rooms without dependence, it is exactly the value of $fet_{i}$ since they are those that ``start" the crowd movement in the estimation. %$maxSpd$  is the maximum speed an agent .
Therefore, $fet_{i}$ and $gfet_{i}$ are used to propagate values to the following rooms in the environment. In addition, $pop$ is the final population considered in a given room, i.e., $pop=ip+(f.F)$, and for $r_k$ without dependent rooms it is set as $pop_k=ip_k$.
\end{enumerate}

% \item {If $D_i<>0$ then Estimate(all $r_{D_i}$)}:

% \subitem {Having needed data in place for $r_{D_i}$ (room width, room length, exit size - three coming from the Environment Editor - Input flow $f_{D_i}$, Flow duration $F_{D_i}$ and initial population $pop_{D_i}$) the estimation of $r_{D_i}$ can happen. The neural networks are loaded and each output parameter ($st_{D_i}$, $\bar{t_{D_i}}$, 
% $\bar{s_{D_i}}$ and 
% $\bar{d_{D_i}}$) is estimated using its respective network. In addition, other features are computed:

% \begin{eqnarray}
%  fet_{D_i} = min(at_{D_i}+fet_{D_i}),
% \label{eq:it}
% \end{eqnarray}

% \begin{eqnarray}
%  it_{D_i} = min(at_{D_i}+fet_{D_i}),
% \label{eq:it}
% \end{eqnarray}
% }

For each room with $N_{D_i}>0$ (see rooms $i$ and $j$ in Figure~\ref{fig:Heuristics}), we perform the following execution flow: first, if the room has a dependence ($r_{D_i}$) that was not yet estimated, the estimation of this room is postponed until all dependents are estimated.

\begin{enumerate}
\item In order to estimate the population in $r_i$, we need to know the flow of people who enter in this room coming from other rooms. In our model, those concepts are represented through the following parameters: $F_i$ (flow duration defined in Equation~\ref{eq:flowduration1}) and $f_i$ (input flow defined in Equation~\ref{eq:flowduration2}), which consider data coming from dependent rooms ($r_{D_i}$) as follows:

%These two variables together determines a window in time in which the dependence rooms of $r_i$ will send agents to $r_i$, as described in Equation~\ref{eq:flowduration1}. The input flow is determined as Equation~\ref{eq:flowduration2}. This way all agents coming into $r_i$ from its dependences are considered a constant flux of agents during the time window.
\begin{eqnarray}
 F_i = (max(gfet_{D_i} - fet_{D_i} + tt_{D_i})) - min(gfet_{D_i}),
 %F_i = ift_i - git_i.
\label{eq:flowduration1}
\end{eqnarray}
\begin{eqnarray}
f_i = \frac{\sum_{i=1}^{N_{D_i}} (pop_{D_i}  .  in_{D_i})}{F_i},
\label{eq:flowduration2}
\end{eqnarray}
where $N_{D_i}$ is the number of dependent rooms of $i$, $in_{D_i}$ is the percentage of agents from $D_i$ that move to $r_i$, and $pop_{D_i}$ is the population of $r_{D_i}$.
%
%\begin{eqnarray}
% git_i = min(gfet_{D_i}), \\
 %ift_i = max(gfet_{D_i} - fet_{D_i} + st_{D_i}),
% ift_i = max(gfet_{D_i} - fet_{D_i} + tt_{D_i}),
%\label{eq:iit}
%\end{eqnarray}

%, and it is computed as follows:
%\begin{eqnarray}
%pop_i = ip_i + (f_i. F_i),
%\label{eq:pop}
%\end{eqnarray}
%where $ip_i$ is the number of agents initially created in room $i$.
%If all ($r_{D_i}$) are already estimated then we compute the following data for $r_i$ (before to estimate such room with ANN)}: %If not, $r_i$ estimation should wait untilthis room estimation should be postponed until its dependences are computed.}
%\item {If all dependencies are already estimated then collect their data to attribute to the \textit{Input flow} $f_r$ and \textit{Flow Duration} $F_r$ input parameters for room $r$, as follows:
%\subitem 
%\begin{eqnarray}
% git_i = min(gfet_{D_i}), \\
 %ift_i = max(gfet_{D_i} - fet_{D_i} + st_{D_i}),
 %ift_i = max(gfet_{D_i} - fet_{D_i} + tt_{D_i}),
%\label{eq:iit}
%\end{eqnarray}

%where $git_i$ is the global initial time of $r_i$, i.e. the time this room receives its first agent coming from a $r_{D_i}$ in a global time among the rooms, and $ift_i$ is the income final time of $r_i$, i.e. the global time the last agent coming from a $r_{D_i}$ enters into $r_i$. 

\item %With all the needed data in place for $r_i$ (room width, room length, exit size - three coming from the Environment Editor - Input flow $f_i$, Flow duration $F_i$ and initial population $pop_i$ - calculated as recently explained) the estimation of $r_i$ can happen. 
With the data from all dependents properly propagated for room $i$,
the neural network can estimate the total evacuation time of such room ($tt_i$) as detailed before. %, $\bar{s_i}$ and $\bar{d_i}$) 

\item Finally, in order to propagate the agents parameters to the following rooms, parameters $fet_i$ and $gfet_i$ are computed according to Equations~\ref{eq:fet2} and~\ref{eq:acctime}, respectively. Such equations are different from  Equations~\ref{eq:fet} and ~\ref{eq:it} because room $i$ has dependents, so $pop$ is considered instead of $ip$.

\begin{eqnarray}
 fet_{k} = 
 \begin{cases}
     0, & pop_{k} = 0 \\
     %\frac{length_k}{maxSpd}, & ip_{k} > 0
     \frac{\frac{length_k}{2}}{MaxSpd}, & pop_k > 0,
 \end{cases}
 \label{eq:fet2}
 \end{eqnarray}
 
\begin{eqnarray}
 %gfet_i = fet_i + git_i.
 gfet_i = fet_i + min(gfet_{D_i}).
\label{eq:acctime}
\end{eqnarray}
%Equation~\ref{eq:acctime} is different from~\ref{eq:it} because in the first case room $i$ has $N_{D_i}<>0$, i.e. this room has dependences. Consequently, agents arrive to this room coming from another one, so the time it takes to the first one to exit has an offset of the time it took for someone to enter $r_i$. In the case of Equation~\ref{eq:it}, $N_{D_i}=0$, so agents start to be simulated in time $=0$.

\end{enumerate}

After we finish the estimation of a certain room $r_i$, this entire process is repeated until all $N$ rooms in the environment are estimated. Note that these simple equations have the unique goal of aggregating the flow of population passing through the rooms to compute the global time. The next section presents how such information is aggregated and used to estimate the entire environment.

%Other equations to aggregate could also be possible, we tested some but at the end these bunch of equations had the better results.

\vspace{0.5cm}
\noindent\textit{B~-~Aggregating per-Room Estimations}
\vspace{0.5cm}

The last step of the environment estimation is to estimate the crowd parameter for the entire environment based on the per-rooms estimations.
%To summarize all the rooms estimations into a single environment estimation
We consider the exit rooms data (type $E$) to estimate the  \textit{evacuation total time} $tt_e$ of a specific environment $e$. Indeed, this is the greatest total evacuation time of $M$ existing exit rooms that represent the exits in the environment, computed as:

\begin{eqnarray}
 %st_e = max(at_{E_e}),  
 tt_e = max(git_{E_e} + tt_{E_e}),  
\label{eq:st}
\end{eqnarray}
where $E_e$ are the exit rooms of the environment $e$, and $tt_e$ is the obtained total time estimation using CrowdEst for environment $e$. 

In order to evaluate the precision of our estimation, we compare $tt_e$ with $st_e$, the latter being the simulation time obtained with our simulator as discussed in Section~\ref{sec:results}.

\subsubsection{Environment editor}
\label{sec:EnvEditor}

\begin{figure}[htb]
\centering
\includegraphics[width=0.6\linewidth]{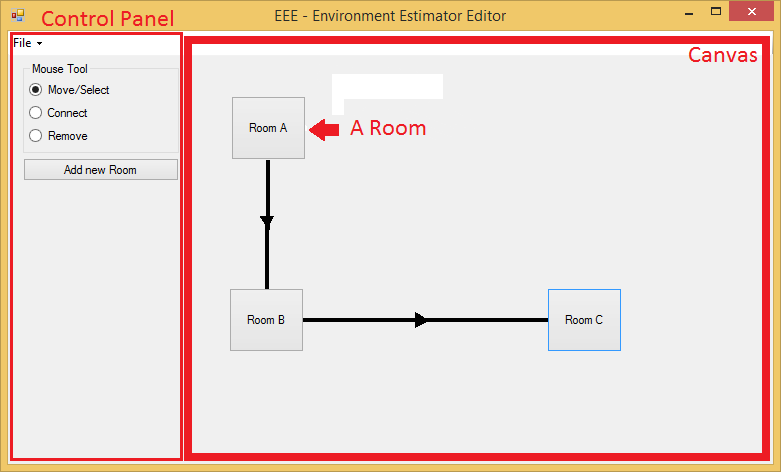}
\caption{Environment editor main window, with the control panel and the canvas regions. Arrows indicate the connections between rooms.}
\label{fig:EEE}
\end{figure}

The Environment editor is a simple graphical application to allow users to easily create and edit environments to be used for estimation. It consists of the main window divided into two regions, the control panel and the canvas, as shown in Figure~\ref{fig:EEE}. The control panel provides a file menu with options to save the current environment into a file, load an environment from a file, and to clear the current environment. Also, it has a list of available tools and a button to add a new room to the current environment.

The editor allows for the following actions: i)~move a room in the canvas in order to locate and organize it as the user wants; ii)~create a connection from one room to another, indicating that the agents from the first room will move towards the second one; and iii)~remove an existing room, together with its connections. 

\begin{figure}[h]
\centering
\includegraphics[width=0.6\linewidth]{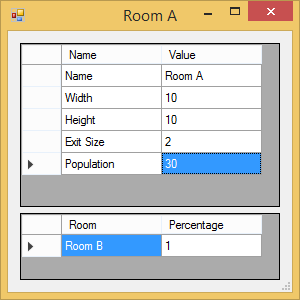}
\caption{Window for configuring room parameters within the editor. It shows the values of the parameters for a given elected room.}
\label{fig:EEE2}
\end{figure}

It is also possible to select a room and check its parameters. In this case, a new window will appear, as in Figure~\ref{fig:EEE2}, showing the \textit{room width}, \textit{room length}, \textit{exit size}, and \textit{initial population} parameters along with the connections presented in the selected room. The value of the parameters can be modified by the user and the connections may be removed at any time. %The remaining input parameters, \textit{Input flow} and \textit{Flow duration}, are not present here because they are calculated during the estimation of the environment, based on the rooms connections and estimation results.

\section{Experimental Results}
\label{sec:results}
This section presents the experimental results. Section~\ref{sec:part1} describes the  results when simulating and estimating 20 full scenarios with variable population distribution within the rooms. %In Section~\ref{sec:Part2} we modeled 30 new environments by replicating rooms and evaluated them in order to measure the impact of ANN error. 
Then, we proceed with a practical usage of our method applying it in a real-life environment and comment about the usability of our application with real users.

% which comprises the usage of ANNs to estimate data in specific rooms, and proposed heuristics to merge the rooms into a complete environment. 
%The tool was developed using the Windows Forms framework, and the application uses a console application in C\#, which receives the file defining the environment by command line.
%As said before, 
%We used a dataset of 18000 rooms for training and 2000 for validating each ANN used.
\subsection{Evaluating Environments with Varied Populations}
\label{sec:part1}

For the first set of scenarios to be tested, we create several distinct environments to be estimated and simulated. Our goal is to show the estimation errors when comparing our method for estimation with the full simulation execution.

\begin{figure*}[!hptb]
\centering

\subfigure[Sim 1]{\includegraphics[scale=0.15]{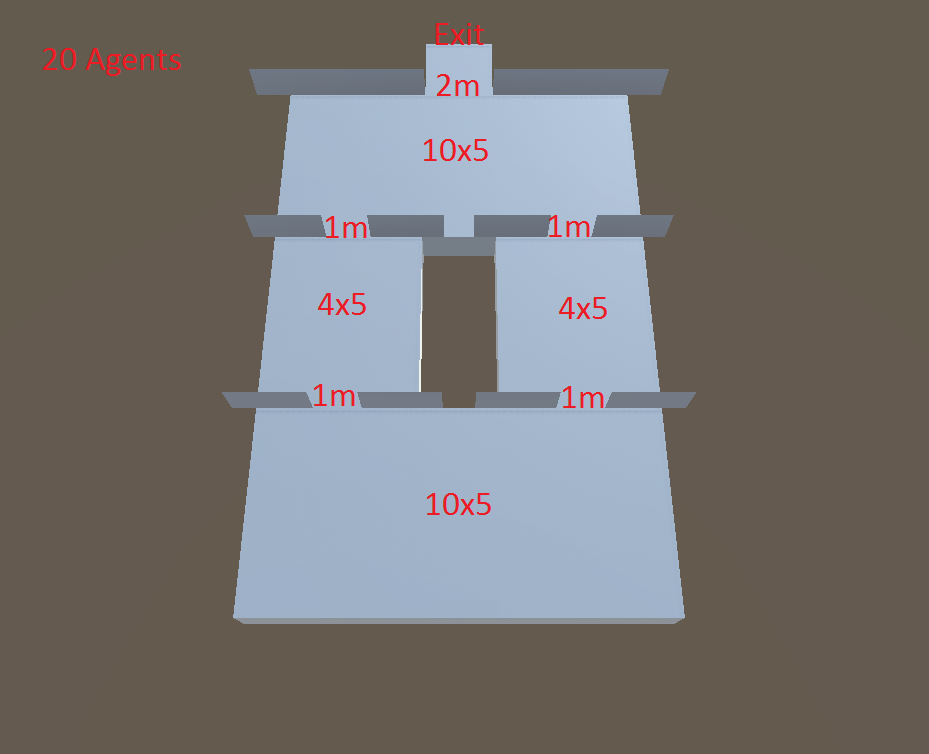}}
\quad
\subfigure[Sim 2]{\includegraphics[scale=0.165]{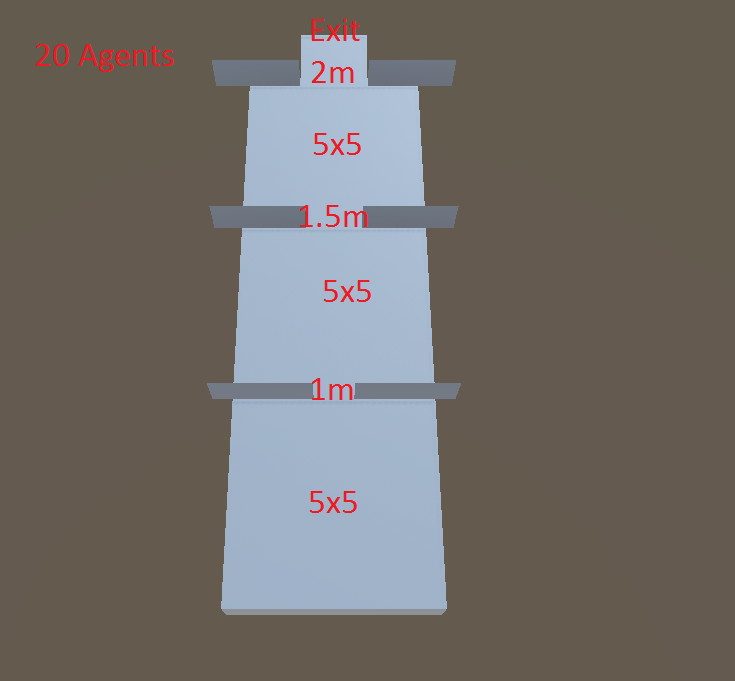}}
\quad
\subfigure[Sim 3]{\includegraphics[scale=0.175]{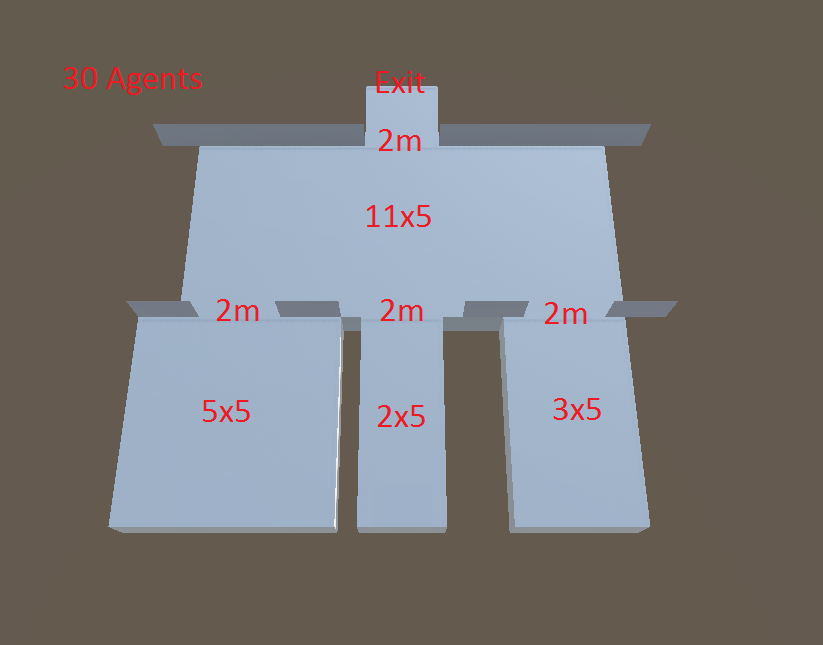}}

\subfigure[Sim 4]{\includegraphics[scale=0.16]{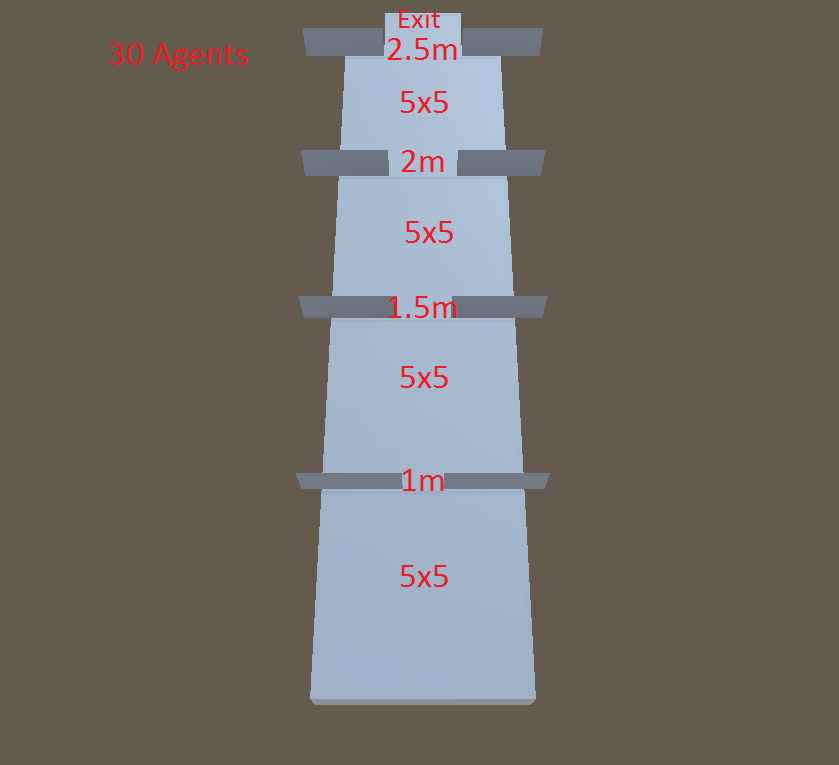}}
\quad
\subfigure[Sim 5]{\includegraphics[scale=0.1]{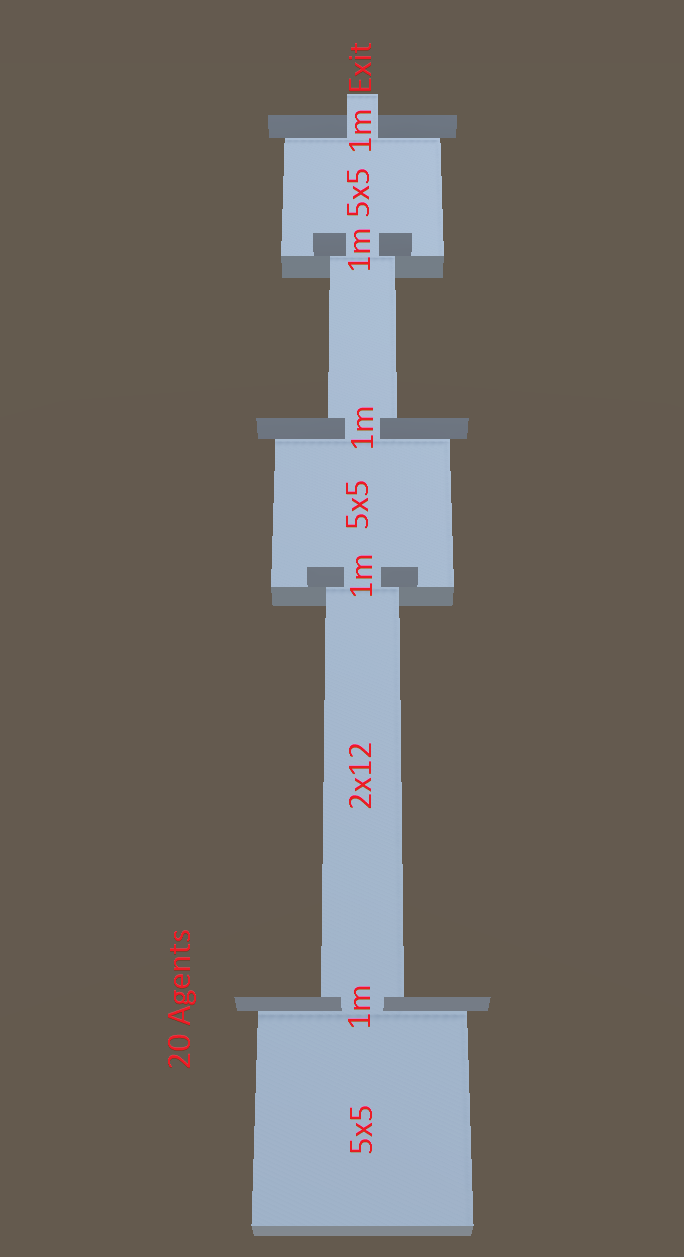}}
\quad
\subfigure[Sim 6]{\includegraphics[scale=0.15]{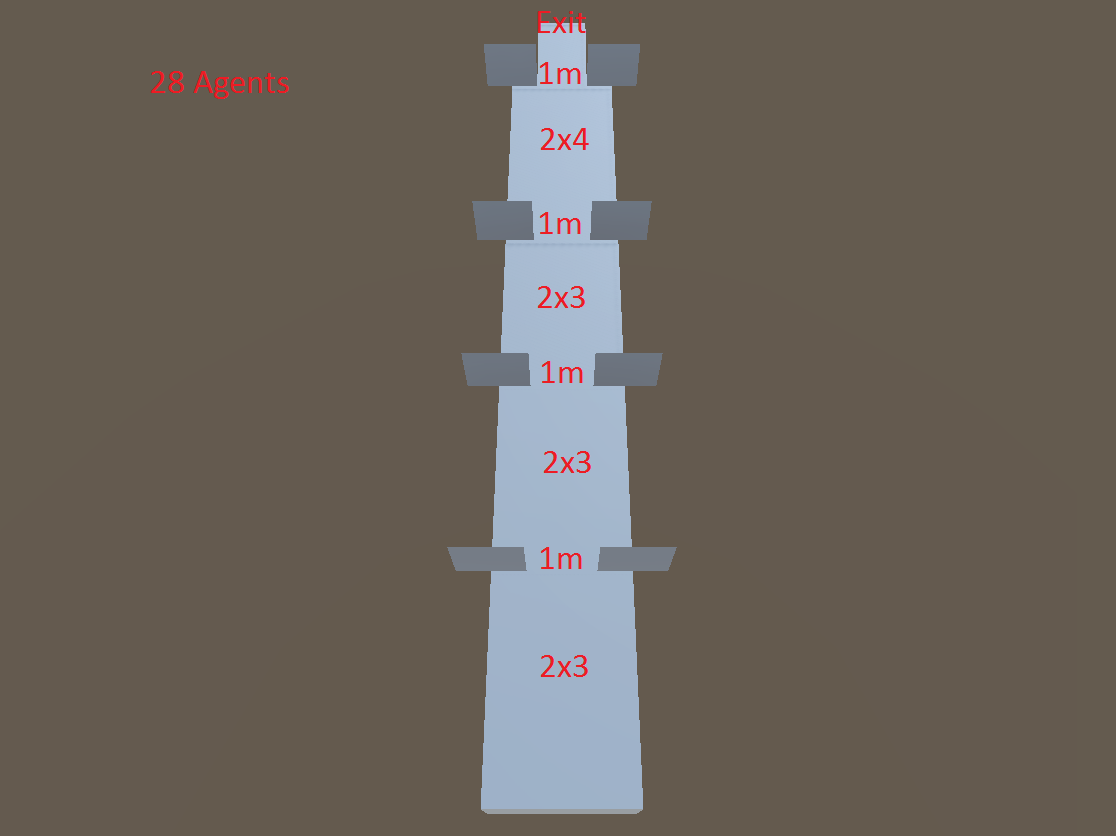}}
\quad
\subfigure[Sim 7]{\includegraphics[scale=0.14]{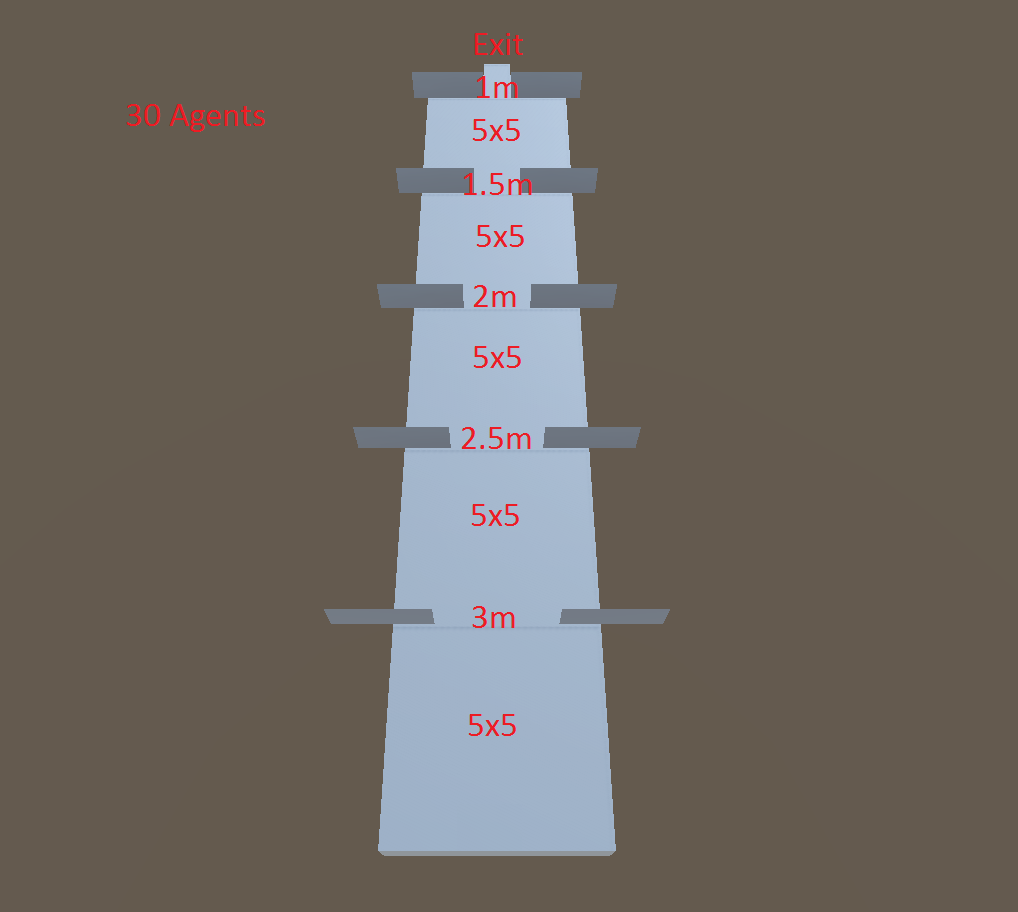}}

\subfigure[Sim 8]{\includegraphics[scale=0.15]{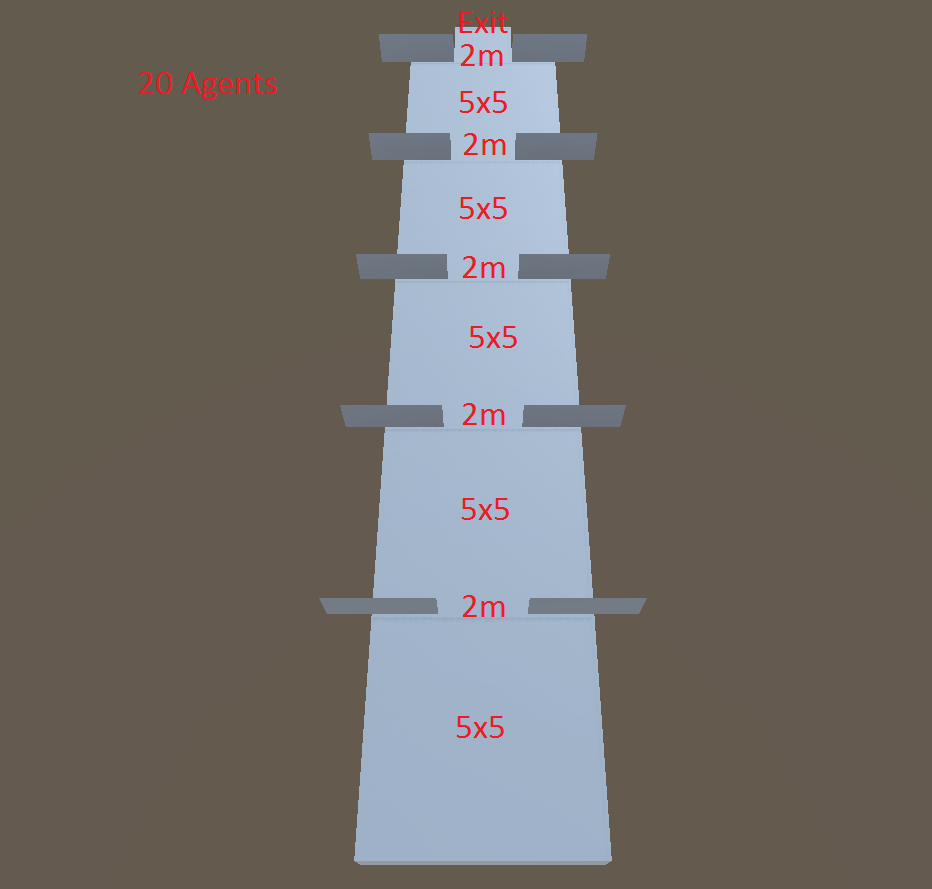}}
\quad
\subfigure[Sim 9]{\includegraphics[scale=0.155]{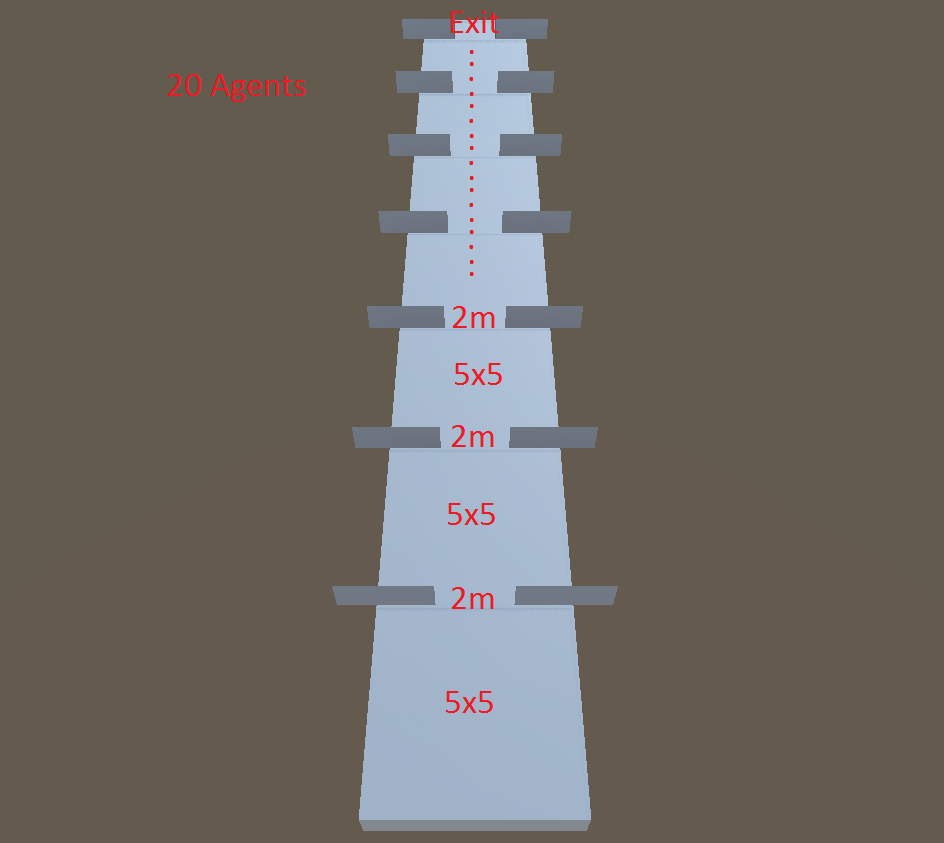}}
\quad
\subfigure[Sim 10]{\includegraphics[scale=0.15]{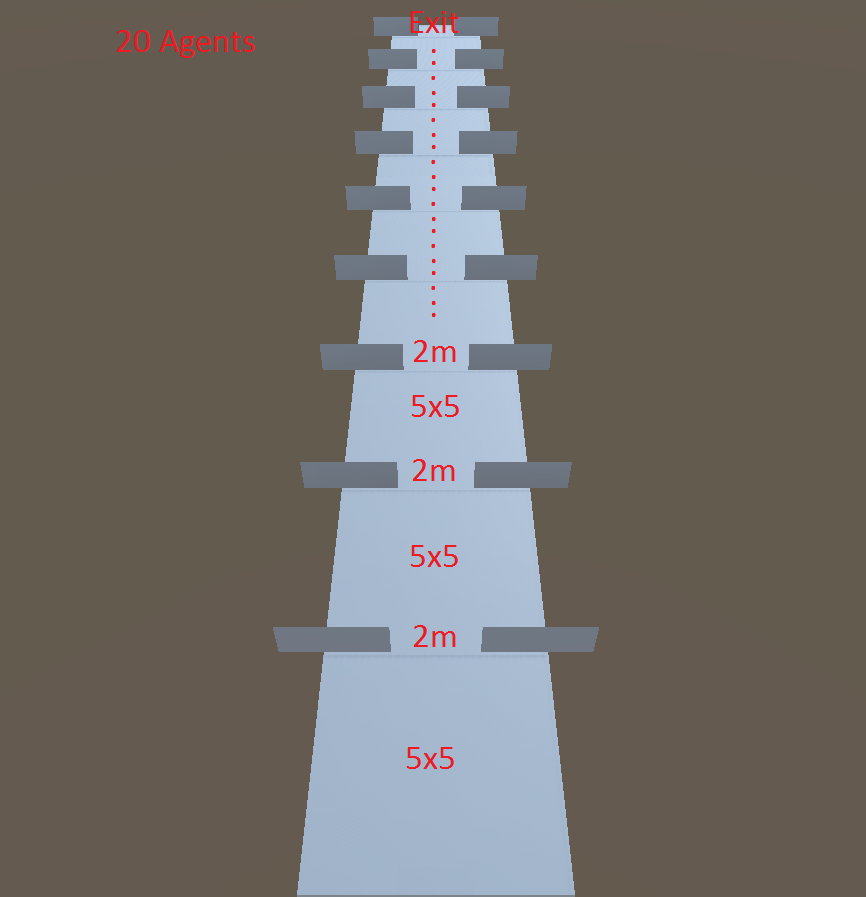}}

\caption{Ten scenarios modeled in the crowd simulator for the 20 cases for evaluating environments. In the first 10 cases, the population is in the first rooms of the environments, while in the second set of simulation, the population is distributed along the rooms.}
\label{fig:salasmodeled}

\end{figure*}

\begin{figure}[!htpb]
\centering
 \subfigure{\includegraphics[width=0.40\linewidth]{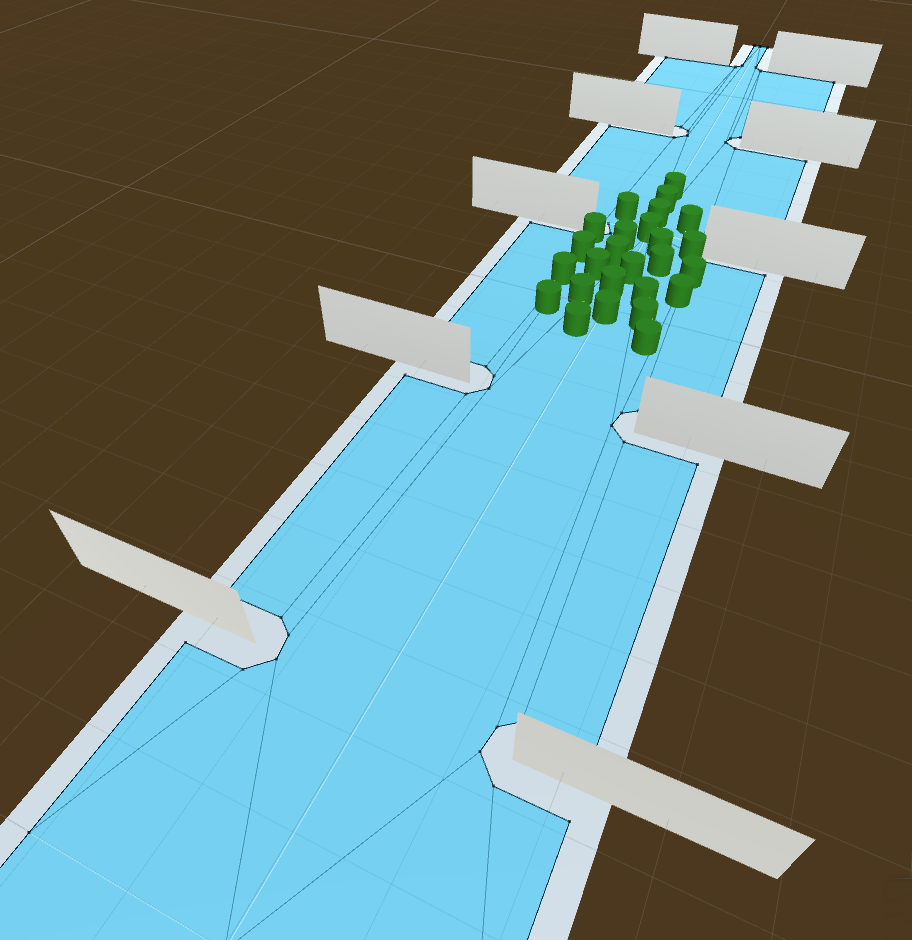}}
 \quad
\subfigure{\fbox{\includegraphics[width=0.52\linewidth]{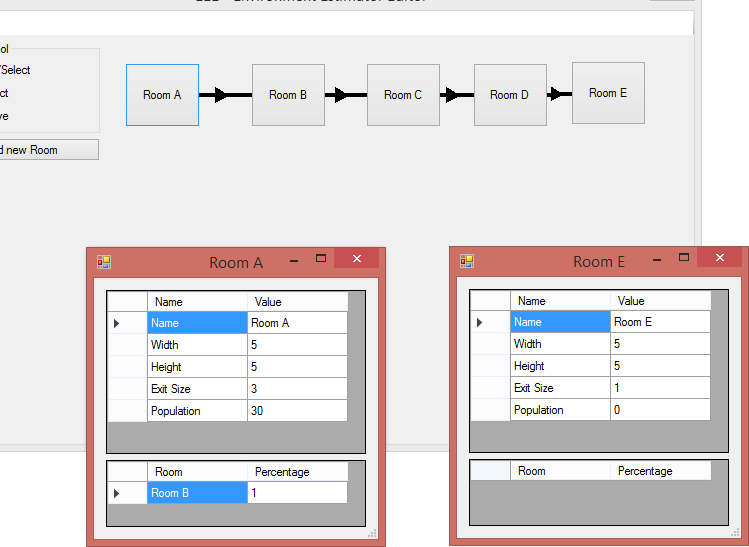}}}

\caption{Example of an environment case: five consecutive rooms with different exit sizes. On the left: the environment at Unity. On the right: the environment at the Environment Editor.}
\label{fig:salas}

\end{figure}

First, using the Environment Editor, we have manually created 10 environments to be tested in our method using the application illustrated in Figure~\ref{fig:EEE}. These environments, illustrated in Figure~\ref{fig:salasmodeled}, comprise a set of consecutive rooms, each of which with its own width, height, and exit sizes. In order to simulate those environments, we used our simulator developed at Unity. Figure~\ref{fig:salas} shows an example of an environment modeled using the Unity Engine (on the left) and using the Environment Editor (on the right).

For each of those environments, we configured two scenarios, one in which the initial population is placed only in the initial rooms of the environment (no dependents) and move towards the exit (last room), and other where the same number of agents is distributed throughout all rooms. %Results are shown in Figure~\ref{fig:err_val}, that contain respectively the percentage relative error of total evacuation time ($tt$) and average time per agent ($\bar{t}$) when compared to the values achieved in the simulations. These 10 simulations were named $Sim_i$, where $i$ is in the interval $[1;10]$.

When analyzing the results, we note that the average errors for the total evacuation time $tt$ for scenarios Sim1 to Sim10 is 20.9\%, while the average error for scenarios Sim1' to Sim10' is 12.6\%. The reason why the environments in which people are only placed in the initial rooms (Sim1 to Sim10) have the largest errors is because crowd simulators can take benefit from the self organization of crowds. A population fixed in the initial rooms of the environment organize themselves in lanes and other crowd structures from the beginning of the simulation, and does not need to change anymore in most of cases. 
CrowdEst, in turn, estimates each room separately without benefiting from the crowd previous organization, so we can say that the worst case when comparing crowd estimation and simulation is when the simulation takes benefit from the crowds' previous organization (Sim1 to Sim10).

\subsection{Testing a Practical Example}

\begin{figure}[htb]
\centering
\includegraphics[width=0.9\linewidth]{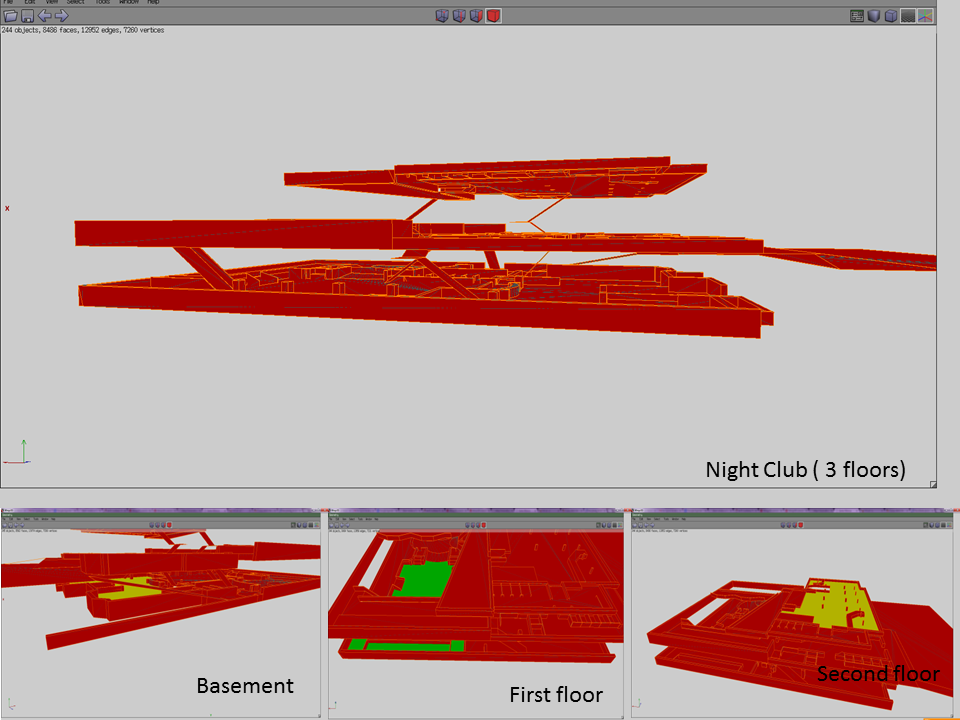}
\caption{Architecture of the nightclub used as base for simulation and estimation.}
\label{fig:SM}
\end{figure}

In order to test our method in a practical example, we use the night club as described in~\cite{Cassol2017}. We used the same 3D environment as illustrated in Figure~\ref{fig:SM} to define the environment to be estimated using our editor and simulate it containing 240 agents, as described in~\cite{Cassol2017}.

\begin{figure}[htb]
\centering
\includegraphics[width=0.9\linewidth]{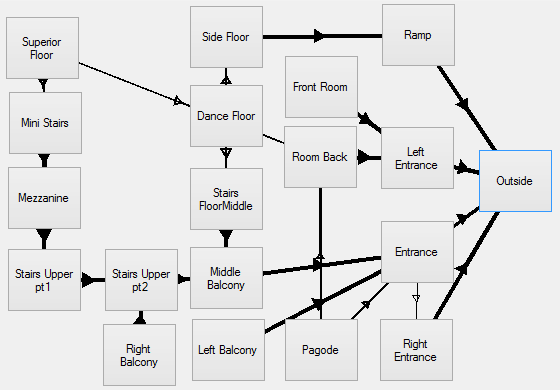}
\caption{Modeling of the SM night club in our editor.} 
\label{fig:SM2}
\end{figure}

The 3D environment has 3 floors and many rooms. There are rooms that have more than one exit (which is a feature that our method does not address), so we proceed by simply summing the widths of the doors to be the exit size value of the room and estimate such room using the ANN. Then, in the heuristic step, we propagate the percentage of population into their correct following rooms, since this information is present in the scenario. This step considers that the crowd is exiting by the doors at the same time and they finish at the same time. Such a decision makes it possible to simulate the night club (and any other environment) using our approach. Nevertheless, it can produce errors and should be better addressed in future work. Figure~\ref{fig:SM2} shows the graph generated using the editor to model the night club (named SM).
%SO: arrumar as palavras que estão em portugues da figura ...
%EST: mais de uma saída até trata. O treinamento sempre foi com uma saida só, mas pra por os números pras salas com várias saídas dá sim. 

%SO: Estevao tens que mudar os nomes das rooms na fig...

% For this practical result, we decided to measure the four metrics initially studied in this work (total evacuation time, average time, average speed and average density). The computed differences are presented in Figure~\ref{fig:SMerror}. As can be seen, the total evacuation time keeps an acceptable error around 30\%, as the average speed and density. On the other hand, the average exit time per agent presents a larger error mainly due to the fact that agents which are placed in each room wait for others who should entry in the specific room to be estimated. This issue was discussed in Figure~\ref{fig:err_val} and it will be the subject of a future investigation.

%\blue{The estimation error of total evacuation time was less than XX\%. Although this number may appear high, it is the result of a comparison with crowd simulations, which already has a distance from reality, and even real evacuation drills lack accuracy when comparing with real evacuation scenarios, table \ref{tab:nightclubtimes} shows the Evacuation time of the night club provided by Cassol et al. \cite{Cassol2017} and the one we obtained in the simulation and estimation we realized in this same scenario. In addition, we imagine that this accuracy nicely compensates the facts that the user do not need to model the 3D environment, learn how to use and prepare the simulation in a crowd simulator, test simulations and finally wait hours for results.}

% \vspace{0.5cm}
% \noindent\textit{A~-~Discussion}
% \vspace{0.5cm}

Table~\ref{tab:nightclubtimes} shows the comparison of evacuation times tested in the night club with several methods. On the left we have the evacuation time obtained in the real drill and simulated  %some CrowdEst estimations with the real drill and simulation 
using CrowdSim (performed by Cassol et al.~\cite{Cassol2017}). %Once we do not know the achieved speed of real people in the scenario (even in the literature~\cite{Cassol2017} it is not mentioned) we tested different possible maximum speeds, altering them on the heuristics Equation \ref{eq:fet} \footnote{This modification helps to estimate scenarios with different speed without needing to do a new training with a database containing different speeds.}, as described in the table. 
On the right we have results obtained with our work, CrowdEst. Note that our estimator  presents an error of approximately +5\% error in comparison to the real-life experiment. % with average speed 0.3m/s, -24\% with 0.4m/s and -39\& with 0.6m/s. We speculate that speeds were not high in real environment, once the night club has many stairs that people should go up and down and they achieved densities of $4.5people/sqm$. In addition, it is known that the crowd simulation time should be smaller than the real life test experiment, once the people do not put themselves in danger/panic, voluntarily. %We also replicate the number of rooms from SM environment, currently has 20 rooms, and generate a new environment with 40 rooms in order to estimate also varying the speeds. Figure 
%SO: não sei se é uma boa explicar isso aqui...
Hence, our simulation method achieves the lowest error when compared with our own estimation method and Cassol's simulator~\cite{Cassol2017}.

 %we estimate the night cub with possible  CrowdEst presents an error of 12\% in relation to the performed Estimation. In addition the difference between our simulator and CrowdEst is 6\%. It suggest that the estimation can be a valuable resource. If we analyze the difference of 12\% (CrowdEst and real experiment) 
It is interesting to see that the previous work using CrowdSim achieves -18\%, and it has been used with experts to guide the crowds in several real-life experiments, so the experts consider such an error to be acceptable in real-life scenarios.
Each simulation/estimation have been executed 10 times and the average results are presented in Table~\ref{tab:nightclubtimes}.

\begin{table}[h]
\centering
\begin{tabular}{|p{1.25cm}|c|c|c|c|c|}
\hline
 & \multicolumn{2}{|c|}{\centering \textbf{Cassol et al.} \cite{Cassol2017}}
 & \multicolumn{2}{|c|}{\centering \textbf{Our work}} \\ \hline
 %& \textbf{Our work}} \\ \hline
 & Real drill & Crowdsim  & CrowdEst  &  OurSimul \\   
 \hline
 Evac time & 175s & 142s & 183 s & 176 s \\ \hline
 Relative error &  & -18.85\% & +4.57\%  & +0.57\% \\ 
 compared with Real drill &  & & &  \\ \hline
\end{tabular}
\quad
\caption{Nightclub evacuation time obtained using different methods and comparison with the real-life drill.}
\label{tab:nightclubtimes}
\end{table}

%SO: Estevao: para simplificar tira speed e density dessa fig...

% \begin{figure}[htb]
% \centering
% \includegraphics[width=1.1\linewidth]{Figs/SMerror}
% \caption{Errors in percentage as resulted from the comparison between the estimation and simulation of a practical example.} 
% \label{fig:SMerror}
% \end{figure}

In terms of advantages when using CrowdEst instead of simulations, we present the following remarks:
\begin{enumerate}
\item Regarding the modeling of the environment:
\subitem - \textbf{5-10 minutes} was the time to model the night club in CrowdEst once the user has the floor plan with the detailed information needed from the environment; 
\subitem - \textbf{12-16 hours} is the time for a skillful designer to model the 3D environment, once s/he has the floor plan of the environment. 

\item Regarding the environment and population configuration:
\subitem - The time to prepare a 3D geometry to be simulated using a crowd simulator is a complex task. The user has to learn how to operate a simulation software, define rooms to walk, place people and groups, define escape routes. The time to accomplish this task using CrowdSim (as defined in~\cite{Cassol2017}) to prepare the night club to be estimated can take \textbf{weeks}, and this needs to be repeated for each new environment to be simulated.
\subitem - Using CrowdEst, we define the population and routes while we define the graph in the editor. \textbf{Ten minutes} was the time to accomplish this task and the task presented in item 1).

\item Regarding the computational time:
\subitem  - The environment estimation is virtually instantaneous, i.e., to estimate the ANN inference time takes \textbf{less than one second} (in CPU, since our method does not require a GPU).
\subitem - As described in~\cite{Cassol2017}, a full simulation containing 240 agents in the night club executing at 24 frames/second takes around \textbf{3 minutes} to evacuate the full environment.
\end{enumerate}

As discussed in~\cite{Cassol2017}, we reinforce that if we want to simulate various distributions of people within the night club, our method can also be used. We can easily define the amount of people in the rooms, e.g., scripting it or using our editor, and instantaneously estimate the total evacuation time. Indeed, considering the night club that has three bifurcations, depending on the granularity used in the population distribution (e.g. 10\% in exit and 90\% in the other) we can have more than 1300 different plans to be executed. As mentioned in~\cite{Cassol2017}, approximately 80 from the 1300 possible plans with 240 agents took about 4 hours, and 1,010 agents took approximately 30 hours to simulate. By using CrowdEst, we can easily and quickly find out the best evacuation plan considering only the evacuation time, among the 1331 plans, which takes approximately 22 mins of processing.

CrowdEst presents an error in comparison to real-life drills, but so does any crowd simulator. The main question is not really to achieve $error=0$, considering that it is impossible even when comparing two different real populations evacuating the same real-life environment. The main intention is to obtain a realistic result not faraway from the expected one. CrowdEst obtains a $5\%$ error in the only available practical example, which is a very promising result. In addition, if such results can be provided by a tool that can be easily tested, we can imagine one application where safety personnel can generate the environment graph, analyze the estimations, and make available the files for the population that is currently in the specific environment. Hence, people can learn and be properly trained for events in an easy and accurate way.

\section{Final Remarks}
\label{sec:remarks}
In this paper, we proposed a methodology for estimating the total evacuation time for complex and generic environments instead of simulating it. We used ANNs to train and estimate data for individual rooms, considering their geometry and population. Then, using such estimations, we propose heuristics to estimate the full environment parameters such as the total evacuation time. The approach we used to train ANNs over individual rooms and then aggregating data using simple heuristics seems quite promising, since we achieve an error of approximately 5\%  when comparing CrowdEst estimations with a real-life drill.

CrowdEst can save weeks (maybe months) of work when compared to the simulation approaches. We believe our tool is a further step in the direction to have all environments from real life estimated and tested. Recall that this work is not about simulation, but estimation, so it explains why our paper does not have any attached videos, as commonly expected in crowd simulation research.

With respect to the limitations of the model: the training dataset was created solely with per-room information, containing only one entrance and one exit directly opposed to the entrance, which may not always be the case in the environments to be tested. We bypassed this problem increasing the width of doors. One way to improve the versatility of the model and to approach its usability for real cases is to add more parameters, which are important to define rooms such as multiple doors and exits. Obstacles are an entire issue by themselves, since there is a lot of information about them: shapes, sizes, positions, orientations, etc, for each obstacle present within the environment. In order to not lose too much of the abstraction of the model, we suggest creating a ``level of obstruction" parameter for the rooms, with a value of 0 meaning the room is free of obstacles and can be traversed without problems, and increasing values for improving the obstacles difficulty on the movement of the agents. Defining this parameter would improve the similarity of the model with reality and we think it is worth a research path of its own.

As future work, we want to test CrowdEst in more real-life environments, besides modeling obstacles and increasing the number of input parameters for refining the ANN estimation.

\section*{Compliance with Ethical Standards}
\begin{itemize}
\item Funding: This work was partially funded by Brazilian Research Agency CNPq (grant number: 305084/2016-0)
\item Disclosure of potential conflicts of interest: Author Est\^ev\~{a}o Testa declares that he has no conflict of interest, author Rodrigo Barros declares that he has no conflict of interest and author Soraia Musse declares that she has no conflict of interest
\item Research involving Human Participants and/or Animals: Not applicable
\item Informed consent: Not applicable
\end{itemize}

% \ifCLASSOPTIONcaptionsoff
%   \newpage
% \fi



%\bibliographystyle{IEEEtran}

\bibliographystyle{spmpsci}
\bibliography{paper}

\end{document}